\documentclass[aps,prd,amsmath,amssymb,twocolumn,twoside,superscriptaddress,
showpacs]{revtex4}

\usepackage{amsfonts}
\usepackage{aas_macros}
\usepackage{graphicx}

\def\hgpc{\ {\rm {\it h}^{-1}Gpc}}
\def\hmpc{\ {\rm {\it h}^{-1}Mpc}}

\def\hmmpc{\ {\rm {\it h}Mpc^{-1}}}

\def\hmsun{\ {\rm M_\odot/{\it h}}}

\def\la{\langle}
\def\ra{\rangle}

\def\dc{\delta_{\rm c}}
\def\dec{\delta_{\rm ec}}

\def\dh{\delta_{\rm h}}
\def\dm{\delta_{\rm m}}
\def\pmm{P_\delta}
\def\pmh{P_{{\rm h}\delta}}
\def\phh{P_{\rm h}}
\def\nh{\bar{n}_{\rm h}}

\def\fnl{f_{\rm NL}^{\rm X}}
\def\floc{f_{\rm NL}^{\rm loc}}
\def\feq{f_{\rm NL}^{\rm eq}}
\def\ffol{f_{\rm NL}^{\rm fol}}
\def\gloc{g_{\rm NL}^{\rm loc}}

\def\vk{{\bf k}}
\def\vq{{\bf q}}
\def\vx{{\bf x}}
\def\vr{{\bf r}}

\def\TM{{\cal M}_R}

\def\apjl{{Astrophys.~ J.~ Lett.~}}

\pacs{98.80.-k,~98.80.Cq}

\begin{document}

\title{Primordial non-Gaussianity in the Large Scale Structure of the Universe}

\author{Vincent Desjacques} \email{dvince@physik.uzh.ch}
\affiliation{Institute for Theoretical Physics, University of Zurich, 
8057 Zurich, Switzerland}
\author{Uro\v s Seljak} \email{seljak@physik.uzh.ch}
\affiliation{Institute for Theoretical Physics, University of Zurich, 
8057 Zurich, Switzerland}
\affiliation{Physics Department, Astronomy Department and Lawrence 
Berkeley National Laboratory, University of California, Berkeley, 
California 94720, USA}

\date{\today}

\begin{abstract}

Primordial non-Gaussianity is a potentially powerful discriminant of
the physical mechanisms that generated the cosmological fluctuations
observed today. Any detection of significant non-Gaussianity would
thus have profound implications for our understanding of cosmic
structure formation. The large scale mass distribution in the Universe
is a sensitive probe of the nature of initial conditions. Recent
theoretical progress together with rapid developments in observational
techniques will enable us to critically confront predictions of
inflationary scenarios and set constraints as competitive as those
from the Cosmic Microwave Background. In this paper, we review past
and current  efforts in the search for  primordial non-Gaussianity in
the large  scale structure of the Universe.

\end{abstract}

\maketitle

\setcounter{footnote}{0}

\section{Introduction}
\label{sec:intro}

In generic inflationary models based on the slow roll of a scalar
field, primordial curvature perturbations are produced by the inflaton
field as it slowly rolls down its potential
\cite{1981JETPL..33..532M,1982PhLB..117..175S,
1982PhLB..115..295H,1982PhRvL..49.1110G}. Most of these scenarios
predict a nearly scale-invariant spectrum of adiabatic curvature
fluctuations, a relatively small amount of gravity waves and tiny
deviations from Gaussianity in the primeval distribution  of curvature
perturbations
\cite{1987PhLB..197...66A,1992PhRvD..46.4232F,1994ApJ...430..447G}.
Although the latest measurements of the cosmic microwave background
(CMB) anisotropies favor a slightly red power spectrum
\cite{2009ApJS..180..330K}, no significant detection of a $B$-mode or
some level of primordial non-Gaussianity (NG) has thus far been
reported from CMB observations.

While the presence of a $B$-mode can only be tested with CMB
measurements, primordial deviations from Gaussianity can leave a
detectable signature in the distribution of CMB anisotropies {\it and}
in the large scale structure (LSS) of the Universe. Until recently, it
was widely accepted that measurement of the CMB furnish the  best
probe of primordial non-Gaussianity \cite{2000MNRAS.313..141V} (see,
e.g., the recent review by E. Komatsu on primordial non-Gaussianity in
the CMB \cite{2010CQGra..27l4010K}).  However, these conclusions  did
not take into account the anomalous scale-dependence of the galaxy
power spectrum and bispectrum arising from primordial NG of the local
$\floc$ type \cite{2004PhRvD..69j3513S,2008PhRvD..77l3514D}. These
theoretical results, together with rapid developments in observational
techniques, will provide large amount of LSS data to critically
confront predictions of non-Gaussian models. In particular, galaxy
clustering should provide independent constraints on the magnitude of
primordial non-Gaussianity as competitive as those from the CMB and,
in the long run, may even give the best constraints.

The purpose of this work is to provide an overview of the search for a
primordial non-Gaussian signal in the large scale structure. We will
begin by briefly summarizing how non-Gaussianity arises in
inflationary models (\S\ref{sec:models}). Next, we will discuss the
impact of primordial non-Gaussianity on the mass distribution in the
low redshift Universe (\S\ref{sec:matterng}). The main body of this
review is \S\ref{sec:lssprobes}, where we describe in detail an number
of methods exploiting the abundance and clustering properties of
observed tracers of the LSS to constrain the amount of initial
non-Gaussianity.  We conclude with a discussion of present and
forecasted constraints achievable with LSS surveys
(\S\ref{sec:limits}).

\section{Models and observables}
\label{sec:models}

Because they assume i) a single dynamical field (the inflaton) ii)
canonical kinetic energy terms (i.e.  perturbations propagate at the
speed of light) iii) slow roll (i.e. the timescale over which the
inflaton field changes is much larger than the Hubble rate) iv) an
initial vacuum state, single-field slow-roll models  lead to a small
level of primordial non-Gaussianity
\cite{1990PhRvD..42.3936S,1992PhRvD..46.4232F,1994ApJ...430..447G}.
The lowest order statistics sensitive to non-Gaussian features in  the
initial distribution of scalar perturbations $\Phi(\vx)$  (we  shall
adopt the standard CMB convention in which $\Phi(\vx)$  is the
Bardeen's curvature perturbation in the matter era) is the 3-point
function or  bispectrum $B_\Phi(\vk_1,\vk_2,\vk_3)$, which is a
function of any triangle $\vk_1+\vk_2+\vk_3=0$ (as follows from
statistical homogeneity which we assume throughout this paper). It has
been recently shown that, in the squeezed limit  $k_3\ll k_1\approx
k_2$, the bispectrum of {\it any} single-field  slow-roll inflationary
model asymptotes to the local shape (defined in
Eq.~{\ref{eq:bphiloc}})
\cite{2003JHEP...05..013M,2003NuPhB.667..119A,2004JCAP...10..006C}.
The corresponding nonlinear parameter predicted by these models is
\begin{equation}
\floc=\frac{5}{12}\left(1-n_s\right)\approx 0.017\qquad\mbox{(single
field)}
\end{equation}
where $n_s$ is the tilt or spectral index of the power spectrum
$P_\Phi(k)$, which is accurately measured to be   $n_s\approx 0.960\pm
0.013$ \cite{2009ApJS..180..330K}. Therefore, any robust measurement
of $\floc$ well above this level would thus rule out single-field
slow-roll inflation.

\subsection{The shape of the primordial bispectrum}

Large, potentially detectable amount of Gaussianity can be produced
when at least one of the assumptions listed above is violated, i.e. by
multiple scalar fields \cite{1997PhRvD..56..535L,2003PhRvD..67b3503L},
nonlinearities in the relation between the primordial scalar
perturbations and  the inflaton field
\cite{1990PhRvD..42.3936S,1994ApJ...430..447G}, interactions of scalar
fields \cite{1993ApJ...403L...1F}, a modified dispersion relation or a
departure from the adiabatic Bunch-Davies ground state
\cite{1997NuPhB.497..479L}. Generation of a large non-Gaussian signal
is also expected at reheating \cite{2004PhRvD..69h3505D} and in the
ekpyrotic scenario  \cite{2008PhRvD..77f3533L}. Each of these physical
mechanisms leaves a distinct signature in the primordial  3-point
function $B_\Phi(\vk_1,\vk_2,\vk_3)$, a measurement of which would
thus provide a wealth of information about the physics of primordial
fluctuations. Although the configuration shape of the primordial
bispectrum can be extremely complex in some models, there are broadly
three classes of shape characterizing the local, equilateral and
folded type of primordial non-Gaussianity
\cite{2004JCAP...08..009B,2009PhRvD..80d3510F}. The magnitude of each
template ``X''  is controlled by a dimensionless nonlinear parameter
$\fnl$ which we seek to constrain using CMB or LSS observations
(instead of attempting a model-independent measurement of $B_\Phi$).

Any non-Gaussianity generated outside the horizon induces a 3-point
function that is peaked on squeezed or collapsed triangles for
realistic values of the scalar spectral index. The resulting
non-Gaussianity depends only on the local value of the Bardeen's
curvature potential, and can thus be conveniently parameterized up to
third order by
\cite{1990PhRvD..42.3936S,1994ApJ...430..447G,2000MNRAS.313..141V,
2001PhRvD..63f3002K}
\begin{equation}
\Phi(\vx)=\phi(\vx)+\floc\phi^2(\vx)+\gloc\phi^3(\vx)\;,
\label{eq:philoc}
\end{equation} 
where $\phi(\vx)$ is an isotropic Gaussian random field and $\floc$,
$\gloc$ are dimensionless, phenomenological parameters. Since
curvature perturbations are of magnitude ${\cal O}(10^{-5})$, the
cubic order correction should always be negligibly small compared to
the quadratic one when ${\cal O}(\floc)\sim{\cal O}(\gloc)$. However,
this condition is not satisfied by some multifield inflationary models
such as the curvaton scenario, in which a large $\gloc$ and a small
$\floc$ can be simultaneously produced \cite{2003PhRvD..67b3503L}.
The quadratic term generates the  3-point function at leading order,
\begin{equation}
\label{eq:bphiloc}
B_\Phi^{\rm loc}(\vk_1,\vk_2,\vk_3)=2\floc\left[P_\phi(k_1)P_\phi(k_2)
+\mbox{(cyc.)}\right]\;,
\end{equation}
where (cyc.) denotes all cyclic permutations of the indices and
$P_\phi(k)$ is the power spectrum of the Gaussian part $\phi(\vx)$ of
the Bardeen  potential. The cubic-order terms generates a trispectrum
$T_\Phi(\vk_1,\vk_2,\vk_3,\vk_4)$ at leading order.

Equilateral type of non-Gaussianity, which arises in inflationary
models with higher-derivative operators such as the DBI model, is well
described by the factorizable form \cite{2006JCAP...05..004C}
\begin{align}
\label{eq:bphieq}
B_\Phi^{\rm eq}(\vk_1,\vk_2,\vk_3) &=6\feq
\Bigl[-\bigl(P_\phi(k_1)P_\phi(k_2)+\mbox{(cyc.)}\bigr)\Bigr.
\nonumber \\ &\quad
\Bigl. -2\bigl(P_\phi(k_1)P_\phi(k_2)P_\phi(k_3)\bigr)^{2/3}\Bigr.  \\
& \quad +  \Bigl. \bigl(P_\phi^{1/3}(k_1)P_\phi^{2/3}(k_2) P_\phi(k_3)
+\mbox{(perm.)}\bigr)\Bigr]\nonumber \;.
\end{align}
It can be easily checked that the signal is largest in the equilateral
configurations $k_1\approx k_2\approx k_3$, and suppressed in the
squeezed limit $k_3\ll k_1\approx k_2$. Note that, in single-field
slow-roll inflation, the 3-point function is a linear combination of
the local and equilateral shape \cite{2003JHEP...05..013M}.

As a third template, we consider the folded or flattened shape  which
is maximized for $k_2\approx k_3\approx k_1/2$
\cite{2009JCAP...05..018M}
\begin{align}
\label{eq:bphifol}
B_\Phi^{\rm fol}(\vk_1,\vk_2,\vk_3) &=6\ffol
\Bigl[\bigl(P_\phi(k_1)P_\phi(k_2)+\mbox{(cyc.)}\bigr)\Bigr.
\nonumber \\ &\quad
\Bigl. +3\bigl(P_\phi(k_1)P_\phi(k_2)P_\phi(k_3)\bigr)^{2/3}\Bigr.  \\
& \quad -  \Bigl. \bigl(P_\phi^{1/3}(k_1)P_\phi^{2/3}(k_2) P_\phi(k_3)
+\mbox{(perm.)}\bigr)\Bigr]\nonumber \;.
\end{align}
and approximate the non-Gaussianity due to modification of the initial
Bunch-Davies vacuum in canonical single field action (the actual 3-point 
function is not factorizable). As in the
previous example, $B_\Phi^{\rm fol}$ is suppressed in the squeezed
configurations. Unlike $B_\Phi^{\rm eq}$ however, the folded shape
induces a scale-dependent bias at large scales
(see~\S\ref{sub:2point}). 

\subsection{Statistics of the linear mass density field}

The Bardeen's curvature potential $\Phi(\vx)$ is  related to the
linear density perturbation  $\delta_0(\vk,z)$  at redshift $z$
through
\begin{equation}
\delta_0(\vk,z)={\cal M}(k,z)\Phi(\vk)\;,
\label{eq:poisson}
\end{equation}
where
\begin{equation}
{\cal M}(k,z)=\frac{2}{3}\frac{k^2 T(k) D(z)}{\Omega_{\rm m} H_0^2}\;.
\label{eq:transfer}
\end{equation}
Here, $T(k)$ is the matter transfer function normalized to unity as
$k\to 0$, $\Omega_{\rm m}$ is the present-day matter density and $D(z)$
is the linear growth rate normalized to $1+z$. $n$-point correlators
of the linear mass density field can thus be written as
\begin{equation}
\la\delta_0(\vk)\cdots\delta_0(\vk_n)\ra= \Biggl(\prod_{i=1}^n{\cal
M}(k_i)\Biggr) \la\Phi(\vk_1)\cdots\Phi(\vk_n)\ra\;.
\end{equation}
Smoothing inevitably arises when comparing observations of the large
scale structure with theoretical predictions from, e.g., perturbation
theory (PT) which are valid only in the weakly nonlinear regime
\cite{2002PhR...367....1B}, or from the spherical collapse model which
ignores the strongly nonlinear internal dynamics  of the collapsing
regions \cite{1972ApJ...176....1G,1996ApJS..103....1B}. For this 
reason, we introduce the {\it smoothed} linear density field
\begin{equation}
\delta_R(\vk,z)={\cal M}(k,z) W_R(k)\Phi(\vk)\equiv\TM(k,z)\Phi(\vk)\;,
\end{equation}
where $W_R(k)$ is a (spherically symmetric) window function of
characteristic radius $R$ or mass scale $M$ that smoothes out the
small-scale nonlinear fluctuations. We will assume a  top-hat  filter
throughout. Furthermore, since $M$ and $R$ are equivalent  variables,
we  shall indistinctly use the notation $\delta_R$ and $\delta_M$ in
what follows.

\subsection{Topological defects models}

In addition to inflationary scenarios, there is a whole class of
models, known as topological defect models, in which cosmological
fluctuations are sourced by an inhomogeneously distributed component
which contributes a small fraction of the total energy momentum tensor
\cite{1999NewAR..43..111D,2000csot.book.....V}. The density field is
obtained as the convolution of a discrete set of points with a
specific density profile.  Defects are deeply rooted in particle
physics as they are expected  to form at a phase transition. Since the
early Universe may have plausibly undergone several phase transitions,
it is rather unlikely that no defects at all were formed. Furthermore,
high redshift tracers of the LSS may be superior to CMB at finding
non-Gaussianity sourced by topological defects
\cite{2001MNRAS.325..412V}. However, CMB observations already provide
stringent limits on the energy density of a defect component
\cite{2009ApJS..180..330K}, so we shall only minimally discuss the
imprint of these scenarios in the large scale
structure. Phenomenological defect models are for instance
\begin{equation}
\delta(\vx)=\phi(\vx)+\alpha_{\rm
NL}\left(\phi^2(\vx)-\la\phi^2\ra\right)
\label{eq:chi2}
\end{equation}
in which the initial matter density $\delta(\vx)$ (rather than the
curvature perturbation $\Phi(\vx)$) contains a term proportional to
the square of a Gaussian scalar field $\phi(\vx)$
\cite{2000MNRAS.313..141V}, or the $\chi^2$ model (also known as
isocurvature CDM model) in which $\delta(\vx)\propto\phi^2(\vx)$
\cite{1999ApJ...510..531P}.

\section{Evolution of the matter density field with primordial NG}
\label{sec:matterng}

In this Section, we summarize a number of results relative to the
effect of primordial NG on the mass density field. These will be
useful to understand the complexity that arises when considering
biased tracers of the density field (see \S\ref{sec:lssprobes}).

\subsection{Setting up non-Gaussian initial conditions}

Investigating the impact of non-Gaussian initial conditions (ICs) on
the large scale structure traced by galaxies etc. requires simulations
large enough so that many long wavelength modes are sampled. At the
same time, the simulations should resolve dark matter halos hosting
luminous red galaxies (LRGs) or quasars (QSOs), so that one can
construct halo samples whose statistical properties mimic as closely
as possible those of the real data. This favors the utilization of
pure N-body simulations for which a large dynamical range can be
achieved.

The evolution of the matter density field with primordial
non-Gaussianity has been studied in series of large cosmological
N-body simulations seeded with Gaussian and non-Gaussian initial
conditions, see e.g.
\cite{1991MNRAS.248..424M,1992MNRAS.259..652W,1993MNRAS.264..749C,
1996ApJ...462L...1G,1999MNRAS.310..511W,2004MNRAS.350..287M,
2008MNRAS.390..438G,2008PhRvD..77l3514D,2009MNRAS.396...85D,
2010MNRAS.402..191P,2009arXiv0911.4768N}.  For generic non-Gaussian
(scalar) random fields, we face the problem of setting up numerical
simulations with a prescribed correlation structure
\cite{2001PASP..113.1009V}. While an implementation of the equilateral
and folded bispectrum shape requires the calculation of several
computationally demanding convolutions, the operation is
straightforward for primordial NG described by a local mapping such as
the $\chi^2$ or the $\floc$ model. In the latter case, the local
transformation $\Phi=\phi+\floc\phi^2$ is performed before
multiplication by the matter transfer function $T(k)$ (computed with
publicly available Boltzmann codes
\cite{1996ApJ...469..437S,2000ApJ...538..473L}).  The (dimensionless)
power spectrum of the Gaussian part $\phi(\vx)$ of the Bardeen
potential is the usual power-law $\Delta_\phi^2(k)\equiv k^3
P_\phi(k)/(2\pi^2)=A_\phi (k/k_0)^{n_s-1}$. Unless otherwise stated,
we shall assume a normalization $A_\phi=7.96\times 10^{-10}$ at the
pivot point $k_0=0.02$Mpc$^{-1}$. To date, essentially all numerical
studies of structure  formation with inflationary non-Gaussianity have
implemented the  local shape solely, so we will focus on this model in
what follows.

Non-Gaussian corrections to the primordial curvature perturbations can
renormalize the input (unrenormalized) power spectrum of fluctuations
used to seed the simulations \cite{2008PhRvD..78l3519M}.  For the
local $\floc$ model with $|\floc|\lesssim 100$, renormalization
effects are unlikely to be noticeable due to the limited dynamical
range of current cosmological simulations.  However, they can be
significant, for example, in simulations of a local cubic coupling
$\gloc\phi^3$  with a large primordial trispectrum
\cite{2010PhRvD..81b3006D}. The cubic order term $\gloc\phi^3$
renormalizes the amplitude $A_\phi$ of the power spectrum of initial
curvature perturbations to $A_\phi\rightarrow
A_\phi+6\gloc\la\phi^2\ra$, where
\begin{equation}
\la\phi^2\ra=\int\!\!\frac{d^3k}{(2\pi)^3}P_\phi(k)\;.
\end{equation}
For scale invariant initial conditions, $\la\phi^2\ra$ has a
logarithmic divergence at large and small scales. In practice however,
the finite box size and the resolution of the simulations naturally
furnish a low- and high-$k$ cutoff. The effective correction to the
amplitude of density fluctuations $\delta\sigma_8$ in the $\gloc\phi^3$
model thus is
\begin{equation}
\delta\sigma_8=3\gloc
\left(\frac{L k_0}{2\pi}\right)^{1-n_s}\left[1-N^{n_s-1}\right]
\frac{A_\phi}{1-n_s}\;,
\label{eq:dsigma8}
\end{equation}
where $N$ is the number of mesh points along one dimension and $L$ is
the simulation box length. For $\gloc=10^6$, $L=1.5\hgpc$ and $N=1024$
for instance, we obtain $\delta\sigma_8\approx 0.015$. 

To generate the initial particle distribution, the Zeldovich
approximation is commonly used instead of the exact gravitational
dynamics. This effectively corresponds to starting from non-Gaussian
initial conditions \cite{1993ApJ...412L...9J}. Since the transition to
the  true gravitational dynamics proceeds rather gradually
\cite{1995MNRAS.274.1049B}, one should ensure that the initial
expansion factor is much smaller than that of the outputs analyzed.
Alternatively, it is possible to generate more accurate ICs based on
second-order Lagrangian perturbation theory (2LPT)
\cite{1998MNRAS.299.1097S}.  At fixed  initial expansion factor, they
reduce transients such that the true dynamics  is recovered more
rapidly \cite{2006MNRAS.373..369C}.

\begin{figure}
\center \resizebox{0.5\textwidth}{!}{\includegraphics{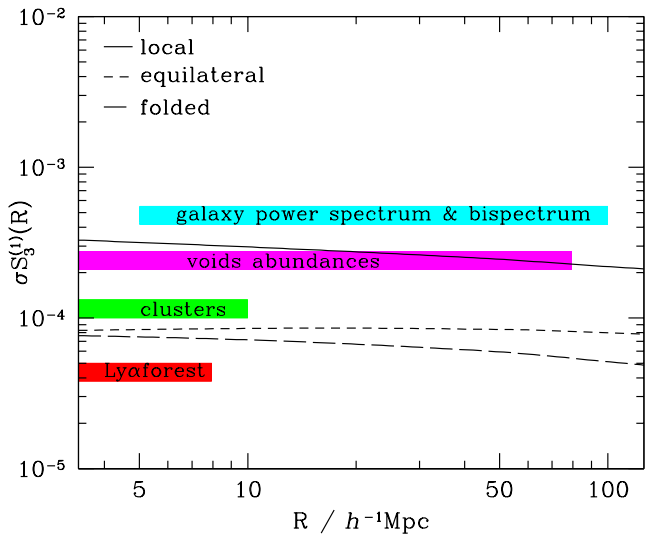}}
\caption{Skewness $\sigma\,S_3^{(1)}\!(R)$ of the smoothed density
field in unit of $\fnl$ for the local, equilateral and folded
bispectrum shape. The skewness for the equilateral and folded
templates is a factor of $\sim 3$ smaller than in the local model. In
any case, this implies that $|\sigma_R S_3(R)|\ll 1$ on the scales
probed  by the large scale structure for realistic values of the
nonlinear coupling parameter, $|\fnl|\lesssim 100$. The shaded regions
approximately indicate the range of scales probed by various LSS
tracers.  For the galaxy power spectrum and bispectrum, the upper
limit sensitively depends upon the surveyed volume.}
\label{fig:skew}
\end{figure}

\subsection{Mass density probability distribution}

In the absence of primordial NG, the probability distribution function
(PDF)  of the initial smoothed density field (the probability
that a randomly placed cell of  volume $V$ has some specific density)
is Gaussian. Namely, all normalized or reduced {\it smoothed}
cumulants $S_J$ of order $J\geq 3$ are  zero. An obvious signature of
primordial NG would thus be an initially non-vanishing skewness
$S_3=\la\delta_R^3\ra_c/\la\delta_R^2\ra^2$ or kurtosis
$S_4=\la\delta_R^4\ra_c/\la\delta_R^2\ra^3-3/\la\delta_R^2\ra$
\cite{1993MNRAS.264..749C,1993ApJ...408...33L,1995MNRAS.274..730L}.
Here, the subscript $c$ denotes the connected piece of the $n$-point
moment that cannot be simplified into a sum over products of lower order
moments. At third order for instance, the cumulant of the smoothed
density field is an integral of the 3-point function,
\begin{equation}
\la\delta_R^3\ra_c=\int\!\!\frac{d^3 k_1}{(2\pi)^3}\int\!\!\frac{d^3 k_2}
{(2\pi)^3}\int\!\!\frac{d^3 k_3}{(2\pi)^3}\,B_R(\vk_1,\vk_2,\vk_3,z)\;,
\end{equation}
where
\begin{align}
B_R(\vk_1,\vk_2,\vk_3,z) &=\TM(k_1,z)\TM(k_2,z)\TM(k_3,z) \nonumber \\
&\quad \times B_\Phi(\vk_1,\vk_2,\vk_3)
\end{align}
is the bispectrum of the smoothed linear density fluctuations at
redshift $z$. Note that, owing to $S_3(R,z)\propto D(z)^{-1}$, the
product $\sigma S_3(R)$ does not depend on redshift. Over the range of
scale accessible to LSS observations, $\sigma S_3(R)$ is a monotonic
decreasing function of $R$ that is of  magnitude  $\sim 10^{-4}$ for
the local, equilateral and folded templates discussed above (Figure
\ref{fig:skew}).   Strictly speaking, all reduced moments should be
specified to fully characterize the density PDF, but a reasonable
description of the density distribution can be achieved with moments
up to the fourth order.

Numerical and analytic studies generally find that a density PDF
initially skewed  towards positive values produces more overdense
regions while a  negatively skewed distribution produces larger voids.
Gravitational instabilities also generate a positive skewness in the
density field, reflecting the fact that the evolved density
distribution exhibits an extended tail towards large overdensities
\cite{1980lssu.book.....P,1984ApJ...279..499F,1991MNRAS.253..727C,
1992ApJ...394L...5B,1993ApJ...412L...9J,1993ApJ...402..387L}. This
gravitationally-induced signal eventually dominates the primordial
contribution such that, at fixed normalization amplitude, non-Gaussian
models deviate more strongly from the Gaussian paradigm at high
redshift. The time evolution of the normalized cumulants $S_J$ can be
worked out for generic Gaussian and non-Gaussian ICs using, e.g., PT
or the spherical collapse approximation. For Gaussian ICs, PT predicts
the normalized cumulants to be time-independent to lowest
non-vanishing order, with a skewness $S_3\approx 34/7$, whereas for
non-Gaussian ICs, the linear contribution to the cumulants decays as
$S_J(R,z)=S_J(R,\infty)/D^{J-2}(z)$
\cite{1994ApJ...429...36F,1998MNRAS.301..524G}.

The  persistence of the primordial hierarchical amplitude
$S_J(R,\infty)$ sensitively depends upon the magnitude of $S_N$,
$N\geq J$, relative to unity. For example, an initially large
non-vanishing kurtosis could source skewness with a time-dependence
and amplitude similar to that induced by nonlinear gravitational
evolution \cite{1994ApJ...429...36F}. Although there is an infinite
class of non-Gaussian models, we can broadly divide them into weakly
and strongly non-Gaussian. In weak NG models, the primeval signal in
the normalized cumulants is rapidly obliterated by gravity-induced
non-Gaussianity. This is the case of hierarchical scaling models where
$n$-point correlation functions satisfy $\xi_n\propto \xi_2^{n-1}$
with $\xi_2\ll 1$ at large scales. By contrast, strongly NG initial
conditions dominate the evolution of the normalized cumulants. This
occurs when the hierarchy of correlation functions obeys the
dimensional scaling $\xi_n\propto \xi_2^{n/2}$, which arises in the
particular case of $\chi^2$ initial conditions
\cite{2000ApJ...542....1S} or in defect models such as texture
\cite{1991PhRvL..66.3093T,1996ApJ...462L...1G,2000PhRvD..62b1301D}.
These scaling laws have been successfully confronted with numerical
investigations of the evolution of cumulants
\cite{1996ApJ...462L...1G,1999MNRAS.310..511W}. We note that the
scaling of the contribution induced by gravity is, however, 
different for the kurtosis \cite{1996MNRAS.279..557C}, suggesting that 
the latter is a better probe of the nature of initial conditions.

Although gravitational clustering tends to erase the memory of initial
conditions, numerical simulations of non-Gaussian initial conditions
show that the occurrence of highly underdense and overdense regions is
very sensitive to the presence of primordial NG. In fact, the  imprint
of primordial NG is best preserved in the negative tail of  the PDF
$P(\rho_R)$ of the evolved (and smoothed) density field $\rho_R$
\cite{2008MNRAS.390..438G}. A satisfactory description  of this effect
can be obtained from an Edgeworth expansion of the initial smoothed
overdensity field \cite{2009MNRAS.395.1743L}. At high densities
$\rho_R\gg 1$, the non-Gaussian modification approximately scales as
$\rho_R^{3/5}$ whereas, at low densities $\rho_R\simeq 0$, the
deviation is steeper, $\rho_R^{6/5}$. Taking into account the weak
scale dependence of  $\sigma S_3(R)$ further enhances this asymmetry.

\begin{figure}
\center \resizebox{0.5\textwidth}{!}{\includegraphics{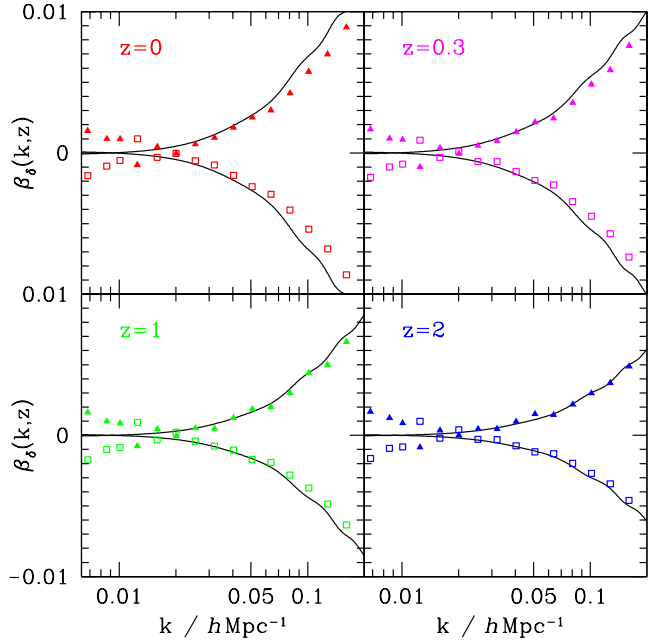}}
\caption{Non-Gaussian fractional correction  $\beta_\delta(k,z)=
\Delta P_\delta^{\rm NG}(k,z)/P_\delta^{\rm G}(k,z)$ to the matter
power spectrum that originates from primordial non-Gaussianity of the
local type.  Results are shown at redshift $z=0$, 0.5, 1 and 2 for
$\floc=+100$ (filled symbols) and $\floc=-100$ (empty symbols). The
solid  curves indicate the prediction from a 1-loop perturbative
expansion.}
\label{fig:psmm}
\end{figure}

\subsection{Power spectrum and bispectrum}

Primordial non-Gaussianity also imprints a signature in Fourier space
statistics of the matter density field.  Positive values of $\floc$
tend to increase the small scale matter power spectrum $\pmm(k)$
\cite{2004PhRvD..69j3513S,2008MNRAS.390..438G,2008PhRvD..78l3534T} and
the large scale matter bispectrum $B_\delta(\vk_1,\vk_2,\vk_3)$
\cite{2004PhRvD..69j3513S,2009PhRvD..80l3002S}.  In the weakly
nonlinear regime where 1-loop PT  applies, the Fourier mode of the
density field for growing-mode initial conditions reads
\cite{1984ApJ...279..499F,1986ApJ...311....6G}
\begin{align}
\label{eq:d1loop}
\delta(\vk,z) &= \delta_0(\vk,z)+\frac{1}{(2\pi)^3} \int\!\!d^3q_1
d^3q_2\,\delta_{\rm D}(\vk-\vq_1-\vq_2) \nonumber \\  & \quad \times
F_2(\vq_1,\vq_2)\delta_0(\vq_1,z)\delta_0(\vq_2,z) \;.
\end{align}
The kernel $F_2(\vk_1,\vk_2)=5/7+\mu(k_1/k_2+k_2/k_1)/2+2\mu^2/7$,
where $\mu$ is the cosine of the angle between $\vk_1$ and $\vk_2$,
describes the nonlinear 2nd order evolution of the density field. It
is nearly independent of $\Omega_{\rm m}$ and $\Omega_\Lambda$ and
vanishes in the (squeezed) limit $\vk_1=-\vk_2$ as a consequence  of
the causality of gravitational instability. At 1-loop PT,
Eq.(\ref{eq:d1loop}) generates the mass power spectrum
\begin{multline}
P_\delta(k,z) = P_\delta^{\rm G}(k,z)+ \Delta P_\delta^{\rm NG}(k,z) 
= P_0(k,z) \\
+\left[P_{(22)}(k,z)+P_{(13)}(k,z)\right]
+\Delta P_\delta^{\rm NG}(k,z)\;.
\end{multline}
Here, $P_0(k,z)$ is the linear matter power spectrum at redshift $z$,
$P_{(22)}$ and $P_{(13)}$ are the standard one-loop contributions in
the case of Gaussian ICs
\cite{1986ApJ...311....6G,1992PhRvD..46..585M},  and
\begin{equation}
\label{eq:p12ng}
\Delta P_\delta^{\rm NG}(k,z)=2\int\!\!\frac{d^3q}{(2\pi)^3}
F_2(\vq,\vk-\vq)B_0(-\vk,\vq,\vk-\vq,z)
\end{equation}
is the correction due to primordial NG \cite{2008PhRvD..78l3534T}.
This terms scales as $\propto D^3(z)$, so the effect of
non-Gaussianity is largest at low redshift. Moreover, because
$F_2(\vk_1,\vk_2)$ vanishes in the squeezed limit, Eq.(\ref{eq:p12ng})
is strongly suppressed at small wavenumbers, even in the local $\floc$
model for which $B_0(-\vk,\vq,\vk-\vq,z)$ is maximized in the same
limit (i.e. $|\vk|\to 0$). For $\floc\sim {\cal O}(10^2)$, the
magnitude of this correction is at a per cent level in the weakly
nonlinear regime $k\lesssim 0.1\hmmpc$
\cite{2009MNRAS.396...85D,2009arXiv0911.4768N,2010PhRvD..81f3530G}, in
good agreement with the measurements (see Figure \ref{fig:psmm}).
Extensions of the renormalization group description of dark matter
clustering \cite{2007JCAP...06..026M} to non-Gaussian initial density
and velocity perturbations can improve the  agreement up to
wavenumbers $k\lesssim 0.25\hmmpc$
\cite{2007PhRvD..76h3517I,2009arXiv0912.4276B}.

It is also instructive to compare measurements of the matter
bispectrum $B_\delta(k_1,k_2,k_3)$ with perturbative predictions.  To
second order in PT, the matter bispectrum is the sum of a primordial
contribution and two terms induced by gravitational instability
\cite{1984ApJ...279..499F,1994ApJ...426...14C} (we will henceforth
omit  the explicit $z$-dependence for brevity),
\begin{align}
\label{eq:bispm}
B_\delta(\vk_1,\vk_2,\vk_3) &= B_0(\vk_1,\vk_2,\vk_3) \\ &\quad 
+\Bigl[2 F_2(\vk_1,\vk_2)P_0(k_1)P_0(k_2)+\mbox{(cyc.)}\Bigr]
\nonumber \\ &\quad 
+\int\!\!\frac{d^3 q}{(2\pi)^3}\Bigl[F_2(\vq,\vk_3-\vq)\Bigr. 
\nonumber \\ &\qquad
\times\Bigl.T_0(\vq,\vk_3-\vq,\vk_1,\vk_2)+\mbox{(cyc.)}\Bigr]
\nonumber \;,
\end{align}
where $T_0(\vk_1,\vk_2,\vk_3,\vk_4)$ is the primordial trispectrum of
the density field. A similar expression can also be derived for
the matter trispectrum, which turns out to be less sensitive to
gravitationally induced nonlinearities \cite{2001ApJ...553...14V}.
The reduced bispectrum $Q_3$ is conveniently defined as
\begin{equation}
Q_3(\vk_1,\vk_2,\vk_3) = \frac{B_\delta(\vk_1,\vk_2,\vk_3)}
{\Bigl[\pmm(k_1)\pmm(k_2)+\mbox{cyclic}\Bigr]}\;.
\end{equation}
For Gaussian initial conditions, $Q_3$ is independent of time and, at
tree-level PT, is constant and equal to $Q_3(k,k,k)=4/7$ for
equilateral configurations \cite{1984ApJ...279..499F}. For general
triangles moreover, it approximately retains this simple behavior,
with a dependence on triangle shape through $F_2(\vk_1,\vk_2)$
\cite{2004PhRvD..69j3513S}. Figure \ref{fig:bisp}  illustrates the
effect of primordial NG of the local $\floc$ type on  shape dependence
of $Q_3$ for a particular set of triangle configurations.   As can be
seen, the inclusion of 1-loop corrections greatly improve the
agreement with the numerical data \cite{2010arXiv1003.0007S}. An
important feature that is not apparent in Fig.\ref{fig:bisp} is the
fact that the primordial part to the reduced matter bispectrum scales
as $Q_3\propto 1/\TM(k)$ for approximately equilateral triangles (and
under the assumption that $\floc$ is scale-independent)
\cite{2004PhRvD..69j3513S}. This anomalous scaling considerably raises
the ability of the matter bispectrum to constrain primordial NG of the
local $\floc$ type. Unfortunately, neither the matter bispectrum  nor
the power spectrum are directly observable with the large scale
structure of the Universe. Temperature anisotropies in the redshifted
21cm background from the pre-reionization epoch could in principle
furnish a direct measurement of these quantities
\cite{2004PhRvL..92u1301L,2005MNRAS.363.1049C,2007ApJ...662....1P},
but foreground contamination may severely hamper any detection.  Weak
lensing is another direct probe of the dark matter, although  we can
only observe it in projection along the line of sight
\cite{2001PhR...340..291B}.

As we will see shortly however, this large-scale anomalous scaling is
also present in the bispectrum and power spectrum of observable
tracers  of the large scale structure such as galaxies. It is this
unique  signature that will make future all-sky LSS surveys
competitive with CMB experiments.

\subsection{Velocities}

Primordial non-Gaussianity also leaves a signature in the large-scale
coherent bulk motions which, in the linear regime, are directly
related to the linear density field \cite{1980lssu.book.....P}.  The
various non-Gaussian models considered by \cite{1995MNRAS.272..859L}
tend to have lower velocity dispersion  and bulk flow than fiducial
Gaussian model, regardless of the sign  of the skewness. However,
while the probability distribution of velocity components is sensitive
to primordial NG of the local type, in defect models it can be  very
close to Gaussian, even when the density field is strongly
non-Gaussian, as a consequence of the central limit theorem
\cite{1992ApJ...390..330S,1995MNRAS.277..927M}. In this regards,
correlation of velocity differences could provide a better test of
non-Gaussian initial conditions \cite{1995ApJ...445....1C}.

To measure peculiar velocities, one must subtract the Hubble flow from
the observed redshift. This requires an estimate of the distance which
is only available for nearby galaxies and clusters (although, e.g.,
the kinetic Sunyaev-Zel'dovich (kSZ) effect could be used to measure
the bulk motions of distant galaxy clusters
\cite{2004ApJ...602...18H}). So far, measurements of the local galaxy
density and velocity fields \cite{1994ApJ...420...44K} as well as
reconstruction of the  initial density PDF from galaxy density and
velocity data \cite{1995ApJ...449..439N} are consistent with Gaussian
initial conditions.

\section{LSS probe of primordial non-Gaussianity}
\label{sec:lssprobes}

Discrete and continuous tracers of the large scale structure such as
galaxies, the Ly$\alpha$ forest, the 21cm hydrogen line etc.,  provide
a distorted image of the matter density field. In CDM cosmologies,
galaxies form inside overdense regions \cite{1978MNRAS.183..341W} and
this introduce a bias between the mass and the galaxy distribution
\cite{1984ApJ...284L...9K}.  As a result, distinct samples of galaxies
trace the matter distribution differently,  the most luminous galaxies
preferentially residing in the most  massive DM halos. This biasing
effect, which concerns most tracers of the large scale structure,
remains to be fully understood. Models of galaxy  clustering assume
for instance that the galaxy biasing relation only depends on the
local mass density, but the actual mapping could be more complex
\cite{2008PhRvD..78j3503D, 2009JCAP...08..020M}. Because of biasing,
tracers of the large scale structure will be affected by primordial
non-Gaussianity in a different way than the mass density field. In
this Section, we describe a number of methods exploiting the abundance
and clustering properties of biased tracers to constrain the level of
primordial NG. We focus on galaxy clustering as it provides the
tightest limits on primordial NG (see \S\ref{sec:limits}).

\subsection{Halo finding algorithm}

Locating groups of bound particles, or DM halos, in simulations is
central to the methods described below. In practice, we aim at
extracting halo catalogs with statistical properties similar to those
of observed galaxies or quasars. This, however, proves to be quite
difficult because the relation  between observed galaxies an DM halos
is somewhat uncertain. Furthermore, there is freedom at defining a
halo  mass.

A important ingredient is the choice of the halo identification
algorithm. Halo finders can be broadly divided into two categories:
Friends-of-Friends (FOF) finders \cite{1985ApJ...292..371D} and
spherical overdensity (SO) finders \cite{1994MNRAS.271..676L}. While
the mass of a SO halo is defined by the radius at which the inner
overdensity exceeds $\Delta_{\rm vir}(z)$ (typically $\sim$ a few
hundred  times the background density $\bar{\rho}(z)$), the mass of a
FOF halo is given by the number of  particles within a linking length
$b$ from  each other ($b\sim 0.15-0.2$ in unit of mean interparticle
distance) from each other. These definitions are somewhat arbitrary
and may suit specific purposes only.  In what follows,  we shall
mainly present results for SO halos as their mass estimate is more
closely connected to  the predictions of the spherical collapse model,
on which most of the analytic formulae presented in this Section are
based. The question of how the spherical overdensity masses can be
mapped onto friends-of-friends masses remains a matter of debate
(e.g. \cite{2009ApJ...692..217L}).  Clearly however, since the peak
height $\nu(M,z)$ depends on the halo mass $M$ through the variance
$\sigma_M$ (see below),  any systematic difference will be reflected
in the value of $\nu$ associated to a specific halo sample.  As we
will see shortly, this affects the size of the fractional deviation
from the Gaussian mass function.

\begin{figure}
\center \resizebox{0.5\textwidth}{!}{\includegraphics{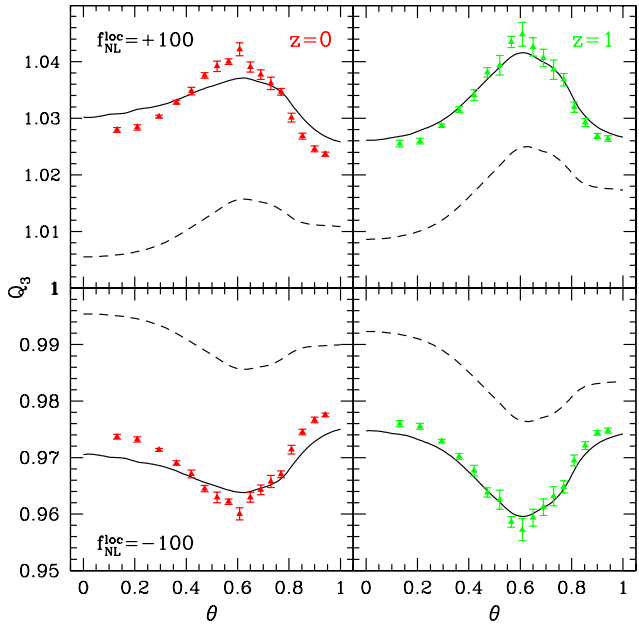}}
\caption{Reduced matter bispectrum $Q_3$ as a function of the angle
$\theta$ between $\vk_1$ and $\vk_2$ for a fixed  $k_1=0.094\hmmpc$
and $k_2=1.5 k_1$. The panels show ratios between the non-Gaussian and
Gaussian $Q_3$ for $\floc=+100$ (top) and $-100$ (bottom). Dashed
lines  correspond to tree-level PT while continuous line indicate the
1-loop PT prediction.}
\label{fig:bisp}
\end{figure}

Catalogs of mock galaxies with luminosities comparable to those of the
targeted survey provide an additional layer of complication that can
be used, among others, to assess the impact of observational errors on
the non-Gaussian signal.  However, most numerical studies of cosmic
structure formation with primordial NG have not yet addressed this
level of sophistication, so we will discuss results based on
statistics of dark matter halos only.

\subsection{Abundances of voids and bound objects}
\label{sub:abundance}

It has long been recognized that departure from Gaussianity can
significantly affect the abundance of highly biased tracers of the
mass density field, as their frequency sensitively depends upon the
tails of the initial density PDF
\cite{1988ApJ...330..535L,1989ApJ...345....3C,1987MNRAS.228..407C}.
The (extended) Press-Schechter approach has been  extensively applied
to ascertain the effect of primordial NG on the high mass tail of the
mass function.

\subsubsection{Press-Schechter approach}

The Press-Schechter theory \cite{1974ApJ...187..425P} and its 
extentions based on excursion sets
\cite{1990MNRAS.243..133P,1991ApJ...367...45C,1991ApJ...379..440B}
predict that the number density $n(M,z)$ of halos of mass $M$ at
redshift $z$ is entirely specified by a  multiplicity function
$f(\nu)$,
\begin{equation} 
n(M,z)=\frac{\bar{\rho}}{M^2}\,\nu f(\nu)\,\frac{d\ln\nu}{d\ln M}\;,
\label{eq:fnu}
\end{equation} 
where the peak height or significance $\nu(M,z)=\dc(z)/\sigma_M$ is
the typical amplitude of fluctuations that produce those halos. Here
and henceforth, $\sigma_M$ denotes the variance of the initial
density field $\delta_M$ smoothed on mass scale $M\propto R^3$ and
linearly extrapolated to present epoch, whereas $\dc(z)\approx 1.68
D(0)/D(z)$ is the critical linear overdensity for (spherical) collapse
at redshift $z$. In the standard Press-Schechter approach, $n(M,z)$ is
related to the level excursion probability $P(>\dc,M)$ that the
linear  density contrast of a region of mass $M$ exceeds $\dc(z)$,
\begin{equation}
\nu f(\nu)= -2 \frac{\bar{\rho}}{M}\,\frac{dP}{dM} = \sqrt{\frac{2}{\pi}}
\nu\,e^{-\nu^2/2}
\end{equation}
where the last equality assumes Gaussian initial conditions. The
factor of 2 is introduced to account for the contribution of low
density regions embedded in overdensities at scale $>M$.  In the
extended Press-Schechter theory, $\delta_M$ evolves with $M$ and
$\nu f(\nu)$ is the probability that a trajectory is absorbed by the
constant barrier $\delta=\dc$ (as is appropriate in the spherical
collapse approximation) on mass scale $M$. In general, the exact form
of $f(\nu)$ depends on the barrier shape \cite{1999MNRAS.308..119S}
and the filter shape \cite{2009arXiv0903.1249M}. Note also that
$\int\!  d\nu\,f(\nu)=1$, which ensures that all the mass is
contained in halos.

\begin{figure}
\center \resizebox{0.5\textwidth}{!}{\includegraphics{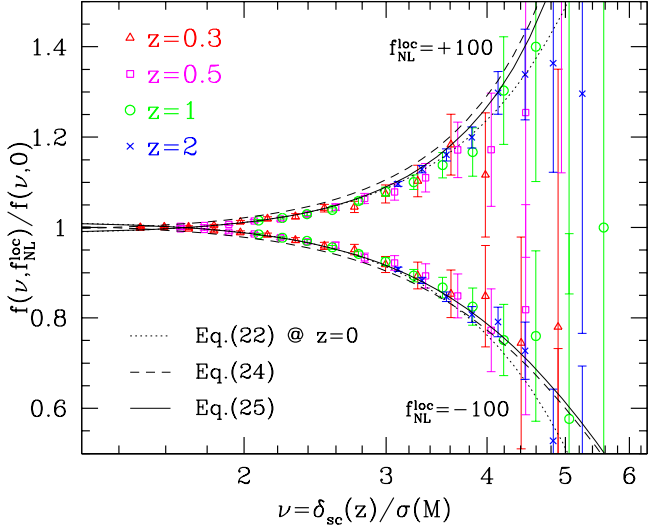}}
\caption{Fractional deviation from the Gaussian mass function as a
function of the peak height $\nu=\dc/\sigma$. Different  symbols refer
to different redshifts as indicated. The various curves are
theoretical prediction at $z=0$ (see text). Halos were identified
using a  spherical overdensity (SO) finder with a redshift-dependent
overdensity  threshold $\Delta_{\rm vir}(z)$ (with $\Delta_{\rm
vir}(z)$ increasing from $\sim 200$ at high redshift to attain $\sim
350$ at $z=0$).  Error bars denote Poisson errors. For illustration,
$M=10^{15}\hmsun$ corresponds to a significance $\nu=3.2$, 5.2, 7.7 at
redshift $z=0$, 1 and 2, respectively. Similarly, $M=10^{14}\hmsun$
and $10^{13}\hmsun$ correspond to $\nu=1.9$, 3, 4.5 and 1.2, 1.9, 2.9
respectively.}
\label{fig:fnu1}
\end{figure}

Despite the fact that the Press-Schechter mass function overpredicts
(underpredicts) the abundance of low (high) mass objects, it can be
used to estimate the fractional deviation from Gaussianity. In this
formalism, the non-Gaussian fractional correction to the multiplicity
function is $R(\nu,\fnl)\equiv
f(\nu,\fnl)/f(\nu,0)=(dP/dM)(>\dc,M,\fnl)/(dP/dM)(>\dc,M,0)$, which is
readily computed  once the non-Gaussian density PDF $P(\delta_M)$ is
known.  In the simple extensions proposed by
\cite{2000ApJ...541...10M} and \cite{2008JCAP...04..014L},
$P(\delta_M)$ is expressed as the inverse transform of a cumulant
generating function.  In \cite{2008JCAP...04..014L}, the saddle-point
technique is applied directly to $P(\delta_M)$. The resulting
Edgeworth expansion is then used to obtain $P(>\dc,M)$. Neglecting
cumulants beyond the skewness, one obtain (we momentarily drop the
subscript $M$ for convenience)
\begin{align}
 R_{_{\rm LV}}(\nu,\fnl) &\approx  
 1+\frac{1}{6}\,\sigma S_3\left(\nu^3-3\nu\right)-
\frac{1}{6}\frac{d\left(\sigma S_3\right)}
{d\ln\nu}\left(\nu-\frac{1}{\nu}\right)
\label{eq:fnulv}
\end{align} 
after integration over regions above  $\dc(z)$. In
\cite{2000ApJ...541...10M}, it is the level excursion probability
$P(>\dc,M)$ that is calculated within the saddle-point approximation.
This approximation better asymptotes to the exact large mass tail,
which exponentially deviates from the Gaussian tail. To enforce the
normalization of the resulting mass function, one may define
$\nu_\star=\delta_\star/\sigma$ with
$\delta_\star=\dc\sqrt{1-S_3\dc/3}$, and use
\cite{2000ApJ...541...10M,2009arXiv0906.1042V}
\begin{equation}
\nu_\star f(\nu_\star)=M^2\,\frac{n_{_{\rm NG}}(M,z)}
{\bar{\rho}}\frac{d\ln M}{d\ln\nu_\star}\;.
\end{equation}
The fractional deviation from the Gaussian mass function then becomes
\begin{equation}
R_{_{\rm MVJ}}(\nu,\fnl)\approx \exp\biggl(\frac{\nu^3}{6}\sigma S_3\biggr)
\Biggl[-\frac{\sigma\nu^2}{6\nu_\star}\frac{d S_3}{d\ln\nu}+
\frac{\nu_\star}{\nu}\Biggr]\;.
\label{eq:fnumvj}
\end{equation}
Both formulae have been shown to give reasonable agreement with
numerical simulations of non-Gaussian cosmologies
\cite{2007MNRAS.382.1261G,2009MNRAS.396...85D,2009MNRAS.398..321G}
(but note that \cite{2007MNRAS.376..343K,2008PhRvD..77l3514D} have
reached somewhat different conclusions).  Expanding $\delta_\star$ at
the first order in $\fnl$ shows that these two theoretical
expectations differ in the coefficient of the $\nu\sigma S_3$
term. Therefore, it is interesting to consider also the approximation
\begin{equation}
\label{eq:fnuthiswork}
R(\nu,\fnl) \approx \exp\biggl(\frac{\nu^3}{6}\sigma S_3\biggr)
\Biggl[1-\frac{\nu}{2}\sigma S_3-\frac{\nu}{6}\frac{d(\sigma S_3)}
{d\ln\nu}\Biggr]\;,
\end{equation}
which is designed to match better the Edgeworth expansion of
\cite{2008JCAP...04..014L} when the peak height is $\nu\sim 1$
\cite{2010PhRvD..81b3006D}.  When the primordial trispectrum is large
(i.e. when $\gloc\sim 10^6$), terms involving the kurtosis must be
included \cite{2000ApJ...541...10M,2008JCAP...04..014L,
2010PhRvD..81b3006D,2009arXiv0910.5125M}. In this case, it is also
important to take into account a possible renormalization of the
fluctuation amplitude, $\sigma_8\to\sigma_8+\delta\sigma_8$
(Eq.\ref{eq:dsigma8}), to which the high mass tail of the mass
function is exponentially sensitive \cite{2010PhRvD..81b3006D}.

\begin{figure}
\center \resizebox{0.5\textwidth}{!}{\includegraphics{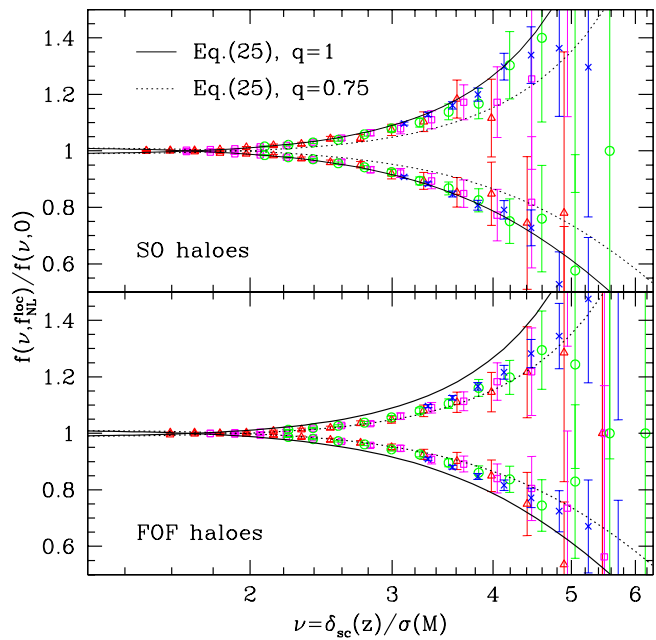}}
\caption{Fractional deviation from the Gaussian mass function as a
function of the peak height $\nu=\dc/\sigma$. Different  symbols refer
to different redshifts as in Fig.~\ref{fig:fnu1}. The curves are the
theoretical prediction Eq.~(\ref{eq:fnuthiswork}) at $z=0$ with $q=1$
(solid) and $q=0.75$ (dotted). In the top panel, halos were identified
using a  spherical overdensity (SO) finder with a redshift-dependent
overdensity  threshold $\Delta_{\rm vir}(z)$  whereas, in the bottom
panel, a  Friends-of-Friends (FOF) finding algorithm with linking
length $b=0.2$ was used.}
\label{fig:fnu2}
\end{figure}

Figure \ref{fig:fnu1} shows the effect of primordial NG of the local
$\floc$ type on the mass function of SO halos identified with a
redshift-dependent overdensity threshold $\Delta_{\rm vir}(z)$
(motivated by spherical collapse in a $\Lambda$CDM cosmology
\cite{1996MNRAS.282..263E}). Overall, the approximation
Eq.(\ref{eq:fnuthiswork}) agrees better with the measurements than
Eq.(\ref{eq:fnumvj}), which somewhat overestimates the data for
$\floc=100$, and than Eq.(\ref{eq:fnulv}), which is not always
positive definite for $\floc=-100$. However, as can be seen in
Fig.~\ref{fig:fnu2}, while the agreement with the data is reasonable
for the SO halos, the theory strongly overestimates the effect in the
mass function of FOF halos.  Reference \cite{2009MNRAS.398..321G}, who
use a FOF algorithm with $b=0.2$, makes the replacement
$\dc\to\dc\sqrt{q}$ with $q\simeq 0.75$ to match the Gaussian and
non-Gaussian mass functions. A physical motivation of  this
replacement is provided by
\cite{2009arXiv0903.1251M,2009arXiv0903.1250M}, who demonstrate that
the diffusive nature of the collapse barrier introduces a similar
factor $q=(1+D_B)^{-1}$ regardless the initial conditions. However,
the value of the diffusion constant $D_B$ was actually measured from
simulations that use a SO finder with constant $\Delta_{\rm vir}=200$
\cite{2009ApJ...696..636R}. In the excursion set approach of
\cite{2002MNRAS.329...61S}, the value of $q$ is obtained by
normalizing the Gaussian mass function to simulation (i.e. it has
nothing to do with the collapse dynamics) and, therefore, depends on
the halo finder. Figure \ref{fig:fnu2} demonstrates that this is also
the case for the non-Gaussian correction $R(\nu,\floc)$: choosing
$q\simeq 0.75$ as advocated in \cite{2009MNRAS.398..321G} gives good
agreement for FOF halos, but strongly underestimates the effect for SO
halos, for which the best-fit $q$ is close to unity. As we will see
below, the strength of the non-Gaussian bias may also be sensitive to
the choice of halo finder.

More sophisticated formulations based on extended Press-Schechter (EPS)
theory and/or modifications of the collapse criterion look promising
since they can reasonably reproduce  both the Gaussian halo counts and
the dependence on $\fnl$
\cite{2009arXiv0903.1250M,2009MNRAS.398.2143L,2009MNRAS.399.1482L}.
The probability of first upcrossing can, in principle, be derived for
any non-Gaussian density field and any choice of smoothing filter
\cite{2000MNRAS.314..354A,2002ApJ...574....9I}.  For a general filter,
the non-Markovian dynamics generates additional terms in the
non-Gaussian correction to the mass function that arise from 3-point
correlators of the  smoothed density $\delta_M$ at different mass
scales \cite{2009arXiv0903.1250M}. However, large error bars still
make difficult to test for the presence of such sub-leading terms.
For generic moving barriers $B(\sigma)$ such as those appearing
in models of triaxial collapse
\cite{2001MNRAS.323....1S,2008MNRAS.388..638D}, the leading
contribution to the non-Gaussian correction approximately is
\cite{2009MNRAS.398.2143L}
\begin{align}
\label{eq:ebarrier}
R(\nu,\fnl) &\approx  1+\frac{1}{6}\,\sigma
S_3\,H_3\Bigl(\frac{B(\sigma)}{\sigma}\Bigr) \\
&\approx 1+\frac{1}{6}\,\sigma S_3\,\sqrt{q}\,(q\nu^3-3\nu)
\nonumber \;,
\end{align}
where $H_3(\nu)\equiv \nu^3-3\nu$ and the last equality assumes
$\nu\gg 1$. For SO halos, Eq.(\ref{eq:ebarrier}) with  $q\sim 0.7$
does not fit to the measured correction $R(\nu,\floc)$ better than
Eq.(\ref{eq:fnuthiswork}). However, the ellipsoidal collapse barrier
with $q\sim 0.7$ matches better the Gaussian mass function for
moderate peak height $\nu\lesssim 2$ \cite{2009MNRAS.399.1482L}.

Parameterizations of the fractional correction based on N-body
simulations have also been considered. While
\cite{2010MNRAS.402..191P} considers a fourth-order polynomial fit to
account for values of  $\floc$ as large as 750,
\cite{2008PhRvD..77l3514D} exploits  the fact that, for sufficiently
small $\floc$, there is a one-to-one mapping between halos in Gaussian
and non-Gaussian cosmologies.  In both cases, the fitting functions
are consistent with  the simulations at the few percent level.

\subsubsection{Clusters abundance}

Rich clusters of galaxies trace the rare, high-density peaks in  the
initial conditions and thus offer the best probe of the high mass tail
of the multiplicity function. To infer the cluster mass function,  the
X-ray and millimeter windows are better suited than the optical-wave
range because selection effects can be understood better (see, however,
\cite{2007astro.ph..3574R}).

Following early theoretical \cite{1986ApJ...310L..21M,
1988ApJ...330..535L,1988PhRvL..61..267C,1989ApJ...345....3C,
1989A&A...215...17B} and numerical
\cite{1991MNRAS.253...35M,1991ApJ...372L..53P,1992MNRAS.259..652W,
1994MNRAS.266..524B} work on the effect of non-Gaussian initial
conditions on the multiplicity function of cosmic structures, the
abundance of clusters and X-ray counts in non-Gaussian cosmologies has
received much attention in the literature. At fixed normalization of
the observed abundance of local clusters, the proto-clusters associated
with high redshift ($2<z<4$) Ly$\alpha$ emitters are much more likely
to develop in strongly non-Gaussian models than in the Gaussian
paradigm
\cite{2004MNRAS.350..287M,2004MNRAS.353..681M,2007MNRAS.376..343K}.
Considering the redshift evolution of cluster abundances thus can
break the degeneracy between the initial density PDF and the
background cosmology.  Simple extensions of  the Press-Schechter
formalism similar to those considered above have been shown to capture
reasonably well the cluster mass function over a wide range of
redshift for various non-Gaussian scenarios
\cite{2000MNRAS.311..781R}.   Scaling relations between the cluster
mass, X-ray temperature and Compton $y$-parameter calibrated using
theory, observations and detailed simulations of cluster formation
\cite{2006ApJ...648..956S,2006ApJ...650..538N}, can be exploited to
predict the observed distribution functions of X-ray and SZ signals
and assess the capability of cluster surveys to test the nature of the
initial conditions
\cite{1998ApJ...494..479C,2002MNRAS.331...71B,2006MNRAS.368.1583S,
2007MNRAS.380..637S,2007ApJ...658..669S,2009MNRAS.397.1125F,
2009arXiv0909.4714R}.

An important limitation of this method is that the impact of realistic
models of primordial non-Gaussianity on cluster abundances is small
compared to systematics in current and upcoming surveys
\cite{1997A&A...320..365O,2000ApJ...534..565H,2004MNRAS.351..375A}.
Given the current uncertainties in the redshift evolution of clusters
(one commonly assumes that clusters are observed at the epoch they
collapse \cite{2000ApJ...534..565H}), the selection effects in the
calibration of X-ray and SZ fluxes with  halo mass, the freedom in the
definition of the halo mass, the degeneracy with the normalization
amplitude $\sigma_8$   (for positive $\fnl$, the mass function is more
enhanced at the high  mass end, and this  is similar to an increase in
the amplitude of  fluctuations $\sigma_8$ \cite{2007JCAP...06..024M})
and the low number statistics, the prospects  of  using the cluster
mass function only to place competitive limits on $\fnl$ with the
current data are small.  A two-fold improvement in cluster mass
calibration is required to provide constraints comparable to CMB
measurements \cite{2004MNRAS.351..375A}.

\subsubsection{Voids abundance}

The frequency of cosmic voids offers a probe of the low density tail
of the initial PDF \cite{2009JCAP...01..010K}. The Press-Schechter
formalism can also be applied to ascertain the sensitivity of the
voids abundance to non-Gaussian initial conditions. Voids are defined
as regions of mass $M$ whose density is less than some critical value
$\delta_v\leq 0$ or, alternatively, whose three eigenvalues of the
tidal tensor \cite{1970Afz.....6..581D} lie below some critical value
$\lambda_v\leq 0$
\cite{2009JCAP...01..010K,2009MNRAS.395.1743L,2009ApJ...701L..25S,
2009MNRAS.399.1482L}. An important aspect in the calculation of the
mass function of voids is the over-counting of voids located inside
collapsing regions. This voids-in-clouds problem, as identified by
\cite{2004MNRAS.350..517S}), can be solved within the excursion set
theory by studying a two barriers problem: $\dc$ for halos and
$\delta_v$ for voids. Including this effect reduces the frequency of
the smallest voids \cite{2009MNRAS.399.1482L}. Neglecting this
complication notwithstanding, the differential number density of voids
of radius $R$ is \cite{2009JCAP...01..010K,2009ApJ...701L..25S}
\begin{equation}
\frac{dn}{dR} \approx \frac{9}{2\pi^2}\sqrt{\frac{\pi}{2}}\,
\frac{|\nu_v|}{R^4}\, e^{-\nu_v^2/2}\, \frac{d\ln|\nu_v|}{d\ln M}
\Biggl[1-\frac{1}{6}\sigma S_3\,H_3\Bigl(|\nu_v|\Bigr)\Biggr]\;,
\end{equation}
where $\nu_v=\delta_v/\sigma_M$.  While a positive $\fnl$ produces
more massive halos, it generates fewer large voids
\cite{2009JCAP...01..010K,2009MNRAS.399.1482L}.  Hence, the effect is
qualitatively different from a simple rescaling of the normalization
amplitude $\sigma_8$. A joint analysis of both abundances of clusters
and cosmic voids might thus provide interesting constraints on the
shape of the primordial 3-point function. There are, however, several
caveats to this method, including the fact that there is no unique way
to define voids \cite{2009JCAP...01..010K}.  Clearly, voids
identification algorithms will have to be tested on numerical
simulations \cite{2008MNRAS.387..933C} before a robust method can be
applied to real data.

\subsection{Galaxy 2-point correlation}
\label{sub:2point}

Before \cite{1984ApJ...284L...9K} showed that, in Gaussian
cosmologies,   correlations  of galaxies and clusters can be amplified
relative to the mass distribution, it was argued that primeval
fluctuations have a non-Gaussian spectrum
\cite{1980ApJ...236..351D,1983ApJ...274....1P} to explain the observed
strong correlation of Abell clusters
\cite{1983ApJ...270...20B,1983SvAL....9...41K}. Along these lines,
\cite{1986ApJ...310...19G} pointed out that  primordial
non-Gaussianity could significantly increase the amplitude  of the
two-point correlation of galaxies and clusters on large scales.
However, except from \cite{2006A&A...457..385A} who showed that
correlations of high density peaks in non-Gaussian models are
significantly stronger than in the Gaussian model with identical mass
power spectrum,  subsequent work focused mostly on abundances
(\S\ref{sub:abundance}) or higher order statistics such as  the
bispectrum  (\S\ref{sub:bispgal}). It is only recently that
\cite{2008PhRvD..77l3514D} have demonstrated the strong
scale-dependent bias arising in non-Gaussian models of the local
$\floc$ type.

\subsubsection{The non-Gaussian bias}

In the original derivation of \cite{2008PhRvD..77l3514D}, the
Laplacian is applied to the local mapping $\Phi=\phi+\floc\phi^2$ in
order to show that, upon substitution of the Poisson equation, the
overdensity in the neighborhood of density peaks is spatially
modulated by a factor proportional to the local value of
$\phi$. Taking into account the coherent motions induced by
gravitational instabilities, the scale-dependent bias correction reads
\begin{equation}
\Delta b_\kappa(k,\floc)= 3\floc \bigl[b_1(M)-1\bigr]\dc(0) 
\frac{\Omega_{\rm m}H_0^2}{k^2 T(k) D(z)}\;,
\label{eq:dbiask}
\end{equation}
where $b_1(M)$ is the linear, Eulerian bias of halos of mass $M$.
This effect can be understood intuitively in terms of a local
rescaling of the small-scale amplitude of matter fluctuations or,
equivalently, a local rescaling of the critical density threshold
\cite{2008PhRvD..77l3514D,2008JCAP...08..031S}.  The original result
missed out a multiplicative factor of $T(k)^{-1}$ which was introduced
subsequently by \cite{2008ApJ...677L..77M} upon a derivation of
Eq.~(\ref{eq:dbiask}) in the limit of high density peaks. The
peak-background split approach
\cite{1986ApJ...304...15B,1989MNRAS.237.1127C,1999MNRAS.308..119S}
promoted by \cite{2008JCAP...08..031S} shows that the scale-dependent
bias applies to any tracer of the matter density field whose
(Gaussian) multiplicity function depends on the local mass density
only. In this approach, the Gaussian piece of the potential is
decomposed into short- and long-wavelength modes,
$\phi=\phi_l+\phi_s$. The short-wavelength piece of the density field
is then given by the convolution
\begin{equation}
\delta_s={\cal M}\star\Phi_s={\cal M}\star\phi_s
\left(1+2\floc\phi_l\right)+\floc\,{\cal M}\star\phi_s^2\;,
\end{equation}
where ${\cal M}$ is the transfer function Eq.(\ref{eq:transfer}).
Ignoring the white-noise term, this provides an intuitive explanation
of the effect in terms of a local rescaling of the small-scale
amplitude of matter fluctuations,
\begin{equation}
\sigma_s\to \sigma_s \Bigl(1+2\floc\phi_l(\vx)\Bigr)\;.
\end{equation}
Assuming the mass function depends only on the peak height
$\nu=\dc/\sigma_s$, the long-wavelength part of the halo overdensity
becomes \cite{2008JCAP...08..031S} (see also
\cite{2008PhRvD..77l3514D, 2008PhRvD..78l3507A,2010PhRvD..81f3530G}
\begin{align}
\delta_l^{\rm h}(\vx) &= \frac{1}{n(\nu)}\,
n\!\left(\frac{\dc-\delta_l(\vx)}
{\sigma_s\left(1+2\floc\phi_l(\vx)\right)}\right)-1 \\ & \approx
-\frac{1}{\sigma_s}
\Bigl(\delta_l(\vx)+2\floc\phi_l(\vx)\Bigr)\frac{d\ln n}{d\nu}
\nonumber\;.
\end{align}
Upon a Fourier transformation and using the fact that, in the Gaussian
case, $\delta_l^{\rm h}(\vk)=b_{\rm L}\delta_l(\vk)$ with the
Lagrangian bias  $b_{\rm L}=-\sigma_s^{-1}d\ln n/d\nu$, we recover the
non-Gaussian  bias correction Eq.(\ref{eq:dbiask}) provided that the
tracers move coherently with the dark matter, i.e. $b_{\rm
L}=b_1(M)-1$ \cite{1996MNRAS.282..347M}.  As emphasized in
\cite{2008PhRvD..77l3514D}, the scale-dependence arises from the fact
that the non-Gaussian curvature perturbation $\Phi(\vx)$ is related to
the  density through the Poisson equation (\ref{eq:poisson}) (so that
$\delta_l(\vk)={\cal M}(k)\phi_l(\vk)$). There is no such effect in
the (local) $\chi^2$ model, Eq.(\ref{eq:chi2}), nor in texture-seeded
cosmologies \cite{1991ApJ...383....1C} for instance.

\begin{figure}
\center \resizebox{0.5\textwidth}{!}{\includegraphics{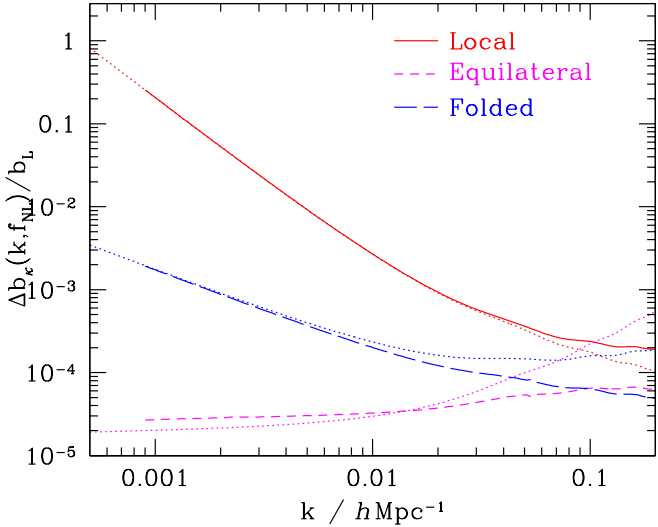}}
\caption{Non-Gaussian bias correction Eq.(\ref{eq:dbfnl}) for the
local, equilateral and folded primordial bispectrum as a function of
wavenumber . Results are shown at $z=0$ for a smoothing radius
$R=5\hmpc$ and a nonlinear parameter $\fnl=1$. The dotted line
represent an analytic  approximation, Eqs (\ref{eq:bapprox}) -- 
(\ref{eq:bapprox2}), which is
valid at large  scales $k\ll 1$. Note that, while the magnitude of
$\Delta b_\kappa(k,\floc)$ does not change with $R$, $\Delta
b_\kappa(k,\feq)$ and $\Delta b_\kappa(k,\ffol)$ strongly depends on
the smoothing radius (see text).}
\label{fig:bfnl}
\end{figure}

The derivation of \cite{2008ApJ...677L..77M}, based on the clustering
of regions of the smoothed density field $\delta_M$ above threshold
$\dc(z)$, is formally valid for  high density peaks
only. However, it is general enough to apply to  any shape of
primordial bispectrum.  The 2-point correlation function of that level
excursion set, which was first derived by \cite{1986ApJ...310L..21M},
can be expressed in the high threshold limit ($\nu\gg 1$) as
\begin{align}
\label{eq:corrlevel}
  \xi_{>\nu}(\vr) &= -1 +
  \exp\Biggl\{\sum_{n=2}^\infty\sum_{j=1}^{n-1}\frac{\nu^n\sigma^{-n}}
  {j!(n-j)!}\Biggr. \\ & \qquad \Biggl. \times
  \xi_R^{(n)}\!\left(\begin{array}{cc}\vx_1,\cdots,\vx_1,  &
  \vx_2,\cdots,\vx_2 \\ j~\mbox{times} & (n-j)~\mbox{times}\end{array},z=0
  \right) \Biggr\}  \nonumber \;,
\end{align}
where $\vr=\vx_1-\vx_2$. For most non-Gaussian models in which the
primordial 3-point function is the dominant correction, this expansion
can be truncated at the third order and Fourier transformed to yield
the non-Gaussian correction $\Delta P_{>\nu}(k)$ to the power
spectrum. Assuming a small level of primordial NG, we can also write
$\Delta P_{>\nu}(k)\approx 2 b_{\rm L}\Delta b_\kappa P_R(k)$ where
$b_{\rm L}\approx \nu^2/\dc$, and eventually obtain
\begin{equation}
\label{eq:dbfnl}
\Delta b_\kappa(k,\fnl)\equiv b_\phi(k){\cal F}(k,\fnl)=
\Biggl(\frac{2 b_{\rm L}\dc(z)}{\TM(k,0)}\Biggr){\cal F}(k,\fnl)\;.
\end{equation}
The dependence on the shape of the 3-point function is encoded in the
function ${\cal F}(k,\fnl)$ \cite{2008ApJ...677L..77M,2009ApJ...706L..91V},
\begin{align}
\label{eq:fk}
{\cal F}(k,\fnl) &=\frac{1}{16\pi^2\sigma^2}\int_0^\infty\!\! 
dk_1\,k_1^2 \TM(k_1,0) \\
&\quad \times\int_{-1}^{+1}\!\!d\mu\,\TM(\sqrt{\alpha},0)
\frac{B_\Phi(k_1,\sqrt{\alpha},k)}{P_\Phi(k)} \nonumber \;,
\end{align}
where $\alpha^2=k^2+k_1^2+2\mu k k_1$. Note that, for $\floc<0$, this
first order approximation always breaks down at sufficiently small $k$
because $\Delta P_{>\nu}(k)<0$.

Figure \ref{fig:bfnl} shows the non-Gaussian halo bias
Eq.(\ref{eq:dbfnl}) induced by the local, equilateral and folded
bispectrum \cite{2009ApJ...706L..91V} .  In the local and folded
non-Gaussianity, the deviation is negligible at $k=0.1\hmmpc$, but
increases rapidly with decreasing wavenumber. Still, the large scale
correction is much smaller for the folded template, and nearly absent
for the equilateral type, which make them much more difficult to
detect with galaxy surveys \cite{2009ApJ...706L..91V}.  To get
insights into the behavior of $\Delta b_\kappa(k,\fnl)$ at large
scales, let us identify the dominant contribution to ${\cal
F}(k,\fnl)$ in the limit $k\ll 1$. Setting
$\TM(\sqrt{\alpha},0)\approx\TM(k_1,0)$  and expanding
$P_\phi(\sqrt{\alpha})$ at second order in $k/k_1$, we find after some
algebra
\begin{align}
\label{eq:bapprox}
{\cal F}(k,\floc) &\approx \floc \\
\label{eq:bapprox1}
{\cal F}(k,\feq) &\approx \feq\,\Bigl[3\,\Sigma_{_{R}}\!
\Bigl(\frac{2(n_s-4)}{3}\Bigr)k^{\frac{2(4-n_s)}{3}}\Bigr. \\
&\qquad \Bigl. + \frac{1}{2}\left(n_s-4\right)\Sigma_{_{R}}\!\left(-2\right)
k^2\Bigr]\,\sigma_R^{-2}\nonumber \\
\label{eq:bapprox2}
{\cal F}(k,\ffol) &\approx \frac{3}{2}\ffol\,
\Sigma_{_{R}}\!\Bigl(\frac{n_s-4}{3}\Bigr)k^{\frac{4-n_s}{3}}\,
\sigma_R^{-2}\;, 
\end{align}
assuming no running scalar index, i.e. $dn_s/d\ln k=0$. The auxiliary 
function $\Sigma_{_{R}}\!(n)$ is defined as
\begin{equation}
\Sigma_{_{R}}\!(n)=\frac{1}{2\pi^2}\int_0^\infty\!\!dk\,k^{(2+n)}\,\TM(k,0)^2
P_\phi(k)\;. 
\end{equation}
Hence, we have $\Sigma_{_{R}}\!(0)\equiv \sigma_R^2$. As can be seen
in Fig.~\ref{fig:bfnl}, these approximations capture relatively well
the large scale non-Gaussian bias correction induced by the
equilateral and folded type of non-Gaussianity.  For a nearly
scale-invariant spectrum $n_s\approx 1$, the effect scales as $\Delta
b_\kappa\propto k$ and  $\Delta b_\kappa\propto$ const.,
respectively. Another important feature of the equilateral and folded
non-Gaussian bias is the dependence on the mass scale $M$ through the
multiplicative factor $\sigma_R^{-2}$. Indeed, choosing $R=1\hmpc$
instead of  $R=5\hmpc$ as done in Fig.~\ref{fig:bfnl} would suppress
the effect by a factor of $\sim 3$. In the high peak limit,
$\sigma_R^{-2}\approx b_{\rm L}/\dc(z)$ which cancels out the
dependence on redshift but enhances the sensitivity to the halo bias,
i.e.  $\Delta b_\kappa\propto b_{\rm L}^2$ for the equilateral and
folded shapes whereas $\Delta b_\kappa\propto b_{\rm L}$ in the local
model.

At this point, it is appropriate to mention a few caveats to these
calculations. Firstly, Eq. (\ref{eq:dbiask}) assumes that the tracers
form after a spherical collapse, which may be a good approximation for
the massive halos only. If one instead considers the ellipsoidal
collapse dynamics, in which the evolution of a perturbation  depends
upon the three eigenvalues of the initial tidal shear, $\dc(0)$ should
be replaced by its ellipsoidal counterparts $\dec(0)$ which  is always
larger than the spherical value \cite{2001MNRAS.323....1S}. In this
model, the scale-dependent bias $\Delta b_\kappa$ is thus enhanced by
a factor $\dec(0)/\dc(0)$
\cite{2008PhRvD..77l3514D,2008PhRvD..78l3507A}. Secondly,
Eq.~(\ref{eq:dbiask})  assumes that the biasing of the surveyed
objects is described by the peak height $\nu$ only or, equivalently,
the hosting halo mass  $M$. However, this may not be true for quasars
whose activity may be triggered by merger of halos
\cite{2003MNRAS.343..692H,2005ApJ...630..705H}. Reference
\cite{2008JCAP...08..031S} used the EPS formalism to estimate the bias
correction $\Delta b_{\rm merger}$ induced by mergers,
\begin{equation}
\Delta b_{\rm merger}=\dc^{-1}\;,
\end{equation}
so the factor $b_1(M)-1$ should be replaced by
$b_1(M)-1-\dc^{-1}\approx b_1(M)-1.6$. The validity of this result
should  be evaluated with cosmological simulations of quasars
formation. In this respect, semi-analytic models of galaxy formation
suggest that merger-triggered objects such as quasars do not cluster
much differently than other tracers of the same mass
\cite{2009arXiv0909.0003B}. However, this does not mean that the same
should hold for the non-Gaussian scale dependent bias. Still, since
the recent merger model is an extreme case it seems likely that the
actual bias correction is $0<\Delta b_{\rm merger}<\dc^{-1}$.
Thirdly, the scale-dependent bias has been derived using the
Newtonian approximation to the Poisson equation, so one may wonder
whether general relativistic (GR) corrections to $\TM(k)^{-1}$ may
suppress the effect on scales comparable to the Hubble
radius. Reference \cite{2005JCAP...10..010B} showed how large scale
primordial NG induced by GR corrections propagates onto small scales
once cosmological perturbations reenter the Hubble radius in the
matter dominated era. This effect generates a scale-dependent bias
comparable, albeit of opposite sign to that induced by local NG
\cite{2009ApJ...706L..91V}. More recently, \cite{2009PhRvD..79l3507W}
argues that there are no GR corrections to the non-Gaussian bias and
that the scaling $\Delta b_\kappa\propto k^{-2}$ applies down to
smallish wavenumbers.

\begin{figure}
\center \resizebox{0.5\textwidth}{!}{\includegraphics{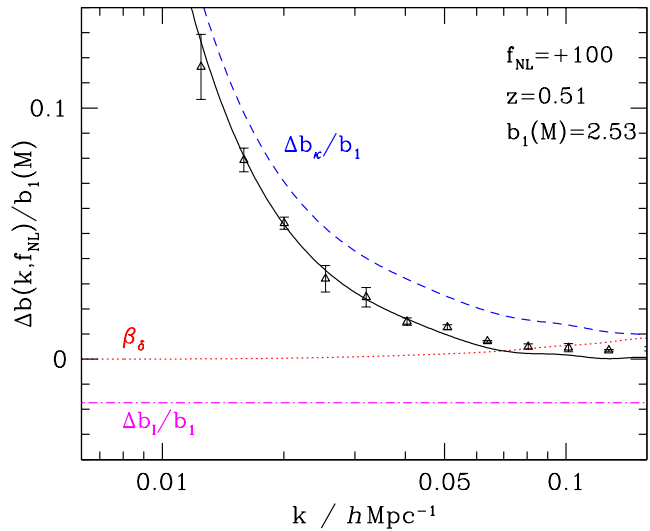}}
\caption{Non-Gaussian bias correction (filled symbols) for halos of
mass $M>2\times 10^{13}\hmsun$ extracted at $z=0.5$ from simulations
of the local $\floc$ model. The solid curve represents the theoretical
model Eq.~(\ref{eq:dbias}). The  dashed, dotted-dashed-dotted and
dotted curves show the scale-dependent bias $\Delta b_\kappa$, the
scale-independent offset $\Delta b_{\rm I}$ and the contribution from
the matter power spectrum $b_1(M)\beta_\delta$ that arise at first
order in $\floc$ (see Fig.\ref{fig:psmm}). The error bars indicate the
scatter among 8 realizations of 1024$^3$ simulations with box size
$L=1600\hmpc$.}
\label{fig:dbias}
\end{figure}

We can also ask ourselves whether  higher-order terms in the series
expansion (\ref{eq:corrlevel})  furnish corrections to the
non-Gaussian bias similar to Eq.(\ref{eq:dbiask}).  The quadratic
coupling $\floc\phi^2$ induces a second order correction to the halo
power spectrum which reads \cite{2010PhRvD..81b3006D}
\begin{align}
\Delta\phh(k)&=\frac{4}{3}(\floc)^2 \Bigl[b_1(M)-1\Bigr]^2
\dc^2(z)\,S_3^{(1)}\!(M) \nonumber \\  &\quad \times\TM(k,0)
P_\phi(k)\;.
\end{align}
Its magnitude relative to the term linear in $\floc$,
Eq.(\ref{eq:dbiask}),  is approximately $0.03$ at redshift $z=1.8$ and
for a halo mass $M=10^{13}\hmsun$.  Although its contribution becomes
increasingly important at higher redshift, it is fairly small for
realistic values of $\floc$. In local NG model, the power spectrum of
biased tracers of the density field can also be obtained from a local
Taylor series in the evolved (Eulerian) density contrast $\delta$ and
the Gaussian part $\phi$ of the initial (Lagrangian) curvature
perturbation \cite{2008PhRvD..78l3519M,2010PhRvD..81f3530G}. Using
this approach, it can be shown that the halo power spectrum arising
from the first order terms of the local bias expansion can be cast
into the form \cite{2008PhRvD..78l3519M}
\begin{equation}
\label{eq:phhloc}
\phh(k)=\bigl[ b_1(M) + \floc b_\phi(k) \bigr]^2 P_R(k)
\end{equation}
Hence, we also obtain a second order term proportional to
$(\floc)^2\TM^{-2}P_R(k)=(\floc)^2P_\phi(k)$ which, however,
contributes only at very small wavenumber $k\lesssim 0.001\hmpc$. All
this suggests that Eq.~(\ref{eq:dbiask}) is the dominant contribution
to the non-Gaussian bias in the wavenumber range $0.001\lesssim
k\lesssim 0.1\hmmpc$.

Finally, a non-Gaussian, scale-dependent bias correction can also
arise in the local, deterministic bias ansatz $\delta_{\rm
h}(\vx)=b_1\delta(\vx)+b_2\delta(\vx)^2/2+\cdots$
\cite{1993ApJ...413..447F} if the initial density field is
non-Gaussian. Here, $b_N$ is the $N$th-order bias parameters (here 
again, the first-order bias is $b_1\equiv 1+b_{\rm L}$). In this
approach, the correction is induced by the correlation $b_1
b_2\la\delta(\vx_1)\delta^2(\vx_2)\ra$ between the linear and
quadratic term in the galaxy biasing relation (which is in fact a
collapsed or squeezed 3-point function) and thus reads
\cite{2008PhRvD..78l3534T,2009PhRvD..80l3002S}
\begin{equation}
\Delta b_\kappa(k,\floc)=2 \floc b_2 \sigma_R^2 {\cal M}_R(k,0)^{-1}\;.
\end{equation}
Even though $b_2\sigma_R^2\approx b_{\rm L}\dc$ in the high-threshold
limit $\nu\gg 1$,  $b_2\sigma_R^2$ behaves very differently than
$b_{\rm L}\dc$ for moderate peak  height because $b_2$ is proportional
to the second derivative of the mass function $n(\nu)$. So far
however, Eq.(\ref{eq:dbiask}) appears to describe reasonably well the
numerical results for a wide range of halo bias.

\subsubsection{Comparison with simulations}

In order to fully exploit the potential of forthcoming large-scale
surveys, a number of studies have tested the theoretical prediction
against the outcome of large numerical simulations
\cite{2008PhRvD..77l3514D,2009MNRAS.396...85D,2010MNRAS.402..191P,
2009MNRAS.398..321G,2009arXiv0911.4768N,2010PhRvD..81f3530G}.

At the lowest order, there are two additional albeit relatively
smaller corrections to the Gaussian bias which arise from the
dependence of both the halo number density $n(M,z)$ and the matter
power spectrum $\pmm(k,z)$ on primordial NG \cite{2009MNRAS.396...85D}.
Firstly, assuming the peak-background split holds, the change in  the
mean number density of halos induces a scale-independent offset which
we denote $\Delta b_{\rm I}(\floc)$.  In terms of the non-Gaussian
fractional correction $R(\nu,\floc)$ to  the mass function, this
contribution is
\begin{equation}
\label{eq:dbiasi}
\Delta b_{\rm I}(\floc) =-\frac{1}{\sigma}\frac{\partial}{\partial\nu}
\ln\Bigl[R(\nu,\floc)\Bigr]\;.
\end{equation}
It is worth noticing that $\Delta b_{\rm I}(\floc)$ has a sign
opposite to that of $\floc$, because the  bias decreases when the mass
function goes up. In practice, the approximation
Eq.~(\ref{eq:fnuthiswork}), which matches well the SO data for
$\nu\lesssim 4$, can be used for moderate values of the peak
height. For FOF halos with linking length $b=0.2$, one should make
the replacement  $\dc\to\dc\sqrt{q}$ with $q\approx 0.75$ in the
calculation of the scale-independent offset. It is sensible to
evaluate $\Delta b_{\rm I}(\floc)$  at a mass scale $\la M\ra$ equal
to the average halo mass of the sample.  Secondly, we also need to
account for the change in the matter power  spectrum (see
Fig.~\ref{fig:psmm} in \S\ref{sec:matterng}). Summarizing, local
non-Gaussianity adds a correction $\Delta b(k,\floc)$  to the bias
$b(k)$ of dark matter halos that reads \cite{2009MNRAS.396...85D}
\begin{equation}
\Delta b(k,\floc)=\Delta b_\kappa(k,\floc)+\Delta b_{\rm I}(\floc)
+b_1(M)\beta_\delta(k,\floc)
\label{eq:dbias}
\end{equation}
at first order in $\floc$. As can be seen in Fig.\ref{fig:dbias}, the
inclusion of these extra terms substantially improves the comparison
between the theory and the simulations.  Considering only the
scale-dependent shift $\Delta b_\kappa$ leads to an apparent
suppression of the effect in simulations relative to the theory.
Including the scale-independent offset $\Delta b_{\rm I}$ considerably
improves the agreement at wavenumbers $k\lesssim 0.05\hmmpc$.
Finally, adding the scale-dependent term $b_1(M)\beta_\delta$ further
adjusts  the match at small scale $k\gtrsim 0.05\hmmpc$ by making the
non-Gaussian  bias shift less negative. Along these lines,
\cite{2010PhRvD..81f3530G} find that the inclusion of  $\Delta b_{\rm
I}$ to the bias also improves the agreement with measurements of
$\Delta b(k,\floc)$ obtained for FOF halos, and show that taking into
account second- and higher-order corrections could extend the validity
of the theory up to scales $k\sim 0.1-0.3\hmmpc$.

\begin{figure}
\center \resizebox{0.5\textwidth}{!}{\includegraphics{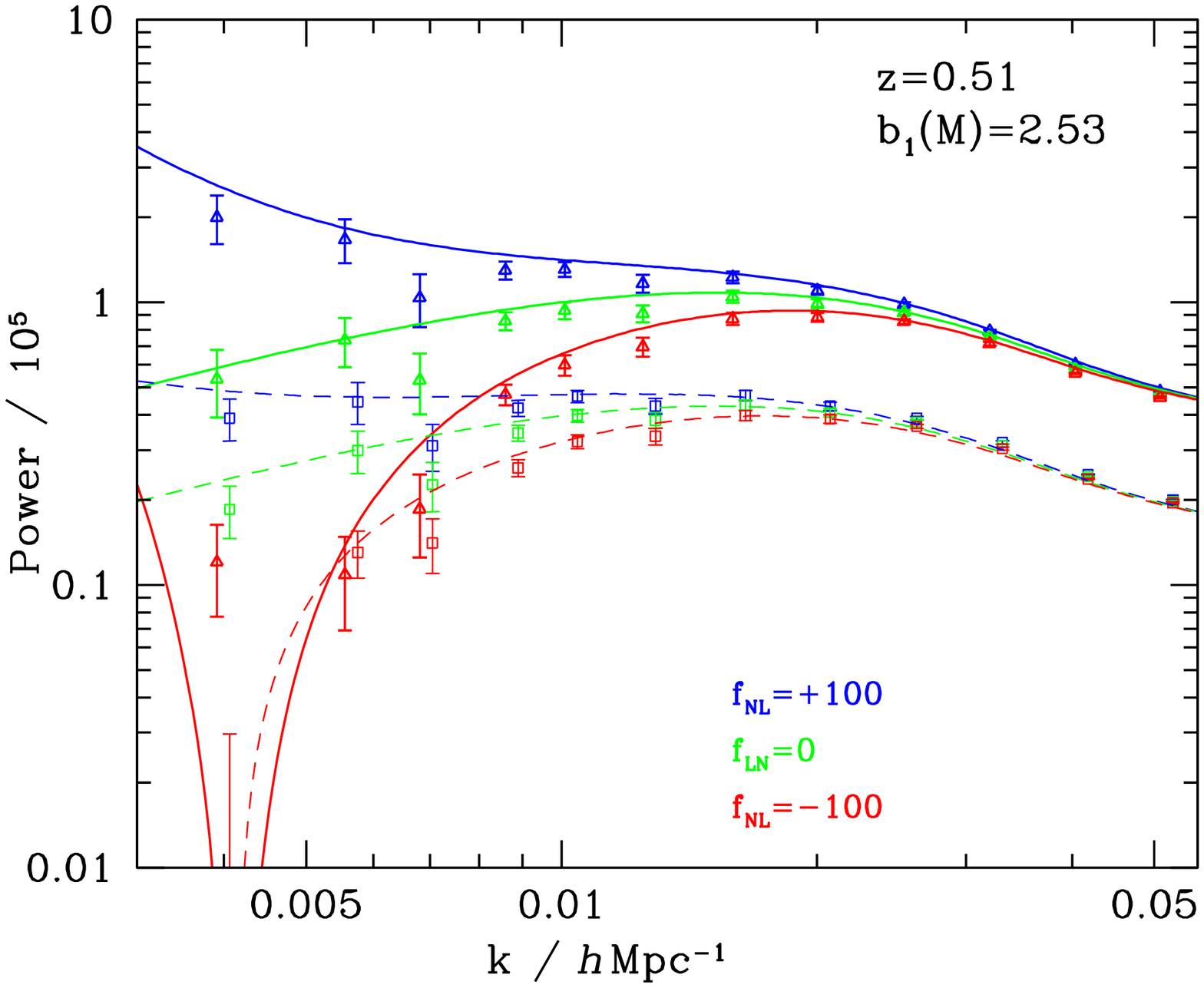}}
\caption{Halo-halo (solid curve) and halo-matter (dashed curve) power
spectra $\phh(k)$ and $\pmh(k)$ measured in simulations of the
Gaussian and $\floc=\pm 100$ models  for halos of mass $M>2\times
10^{13}\hmsun$ at redshift $z=1$.  The error bars represent the
scatter among 8 realizations.  For $\floc=-100$, the cross-power
spectrum is negative on scales $k\lesssim 0.005\hmmpc$, in good
agreement with the theoretical prediction.}
\label{fig:powng}
\end{figure}

The non-Gaussian bias correction can be measured in the cross- and
auto-power spectrum of dark matter halos, $\pmh(k)$ and $\phh(k)$.  To
compute these quantities, dark matter particles and halo centers are
interpolated onto a regular cubical mesh. The resulting dark matter
and halo fluctuation fields, $\dm(\vk)$  and $\dh(\vk)$, are then
Fourier transformed to yield the matter-matter,  halo-matter and
halo-halo power spectra $\pmm(k)$, $\pmh(k)$ and  $\phh(k)$,
respectively. $\phh(k)$ is then corrected for the shot noise, which is
assumed to be  $1/\nh$ if dark matter  halos are a Poisson sampling of
some continuous  field. This discreteness correction is negligible for
$\pmm(k)$ due to the large number of dark matter particles.  On linear
scales ($k\lesssim 0.01\hmmpc$), the halo bias
$b(\vk)=\dh(\vk)/\dm(\vk)$ approaches the constant value $b_1(M)$
which needs to be measured accurately as it  controls the strength of
the scale-dependent bias correction $\Delta b_\kappa$. In this
respect, the ratio $\pmh(k)/\pmm(k)$ is a better proxy for the halo
bias since it is less sensitive to shot-noise.

Auto- and cross-power analyses may not agree with each other if the
halos and dark matter do not trace each other on scale  $k\lesssim
0.01\hmmpc$ where the non-Gaussian bias is large,  i.e. if there is
stochasticity.  Fig.\ref{fig:powng} shows $\pmh(k)$ and $\phh(k)$
averaged over 8 realizations of the models with $\floc=0,\pm 100$
\cite{2009MNRAS.396...85D}. The same Gaussian random seed field $\phi$
was used in each set of runs so as to minimize the sampling
variance. Measurements of the non-Gaussian bias correction  obtained
with the halo-halo or the halo-matter power spectrum are in a good
agreement with each other, indicating that non-Gaussianity does not
induce  stochasticity and the predicted scaling Eq.(\ref{eq:dbiask})
applies equally  well for the auto- and cross-power spectrum. However,
while a number of numerical studies of the $\floc$ model have
confirmed the scaling  $\Delta b_\kappa(k,\floc)\propto\TM(k)^{-1}$
and the redshift dependence $\propto D(z)^{-1}$
\cite{2008PhRvD..77l3514D,2009MNRAS.396...85D,2010MNRAS.402..191P,
2009MNRAS.398..321G}, the exact amplitude of the non-Gaussian bias
correction remains somewhat debatable. Reference
\cite{2009MNRAS.396...85D} who use SO halos and
\cite{2010PhRvD..81f3530G}  who use FOF halos find satisfactory
agreement with the theory once  the scale-independent offset  $\Delta
b_{\rm I}$ is included.  By contrast, \cite{2010MNRAS.402..191P}, who
use the same FOF halos as \cite{2010PhRvD..81f3530G}, argue that the
scale-dependent piece $\Delta b_\kappa$ requires, among others, a
multiplicative  correction of the form $(1-\beta_1\floc)$, with
$\beta_1\sim 4\times 10^{-4}>0$.  Similarly,
\cite{2009MNRAS.398..321G} who also use FOF halos find that the theory
is a good fit to the simulations only upon replacing $b_{\rm L}$ by $q
b_{\rm L}$ in Eq.(\ref{eq:dbfnl}), with $q\simeq 0.75$. Part of the
discrepancy may be probably due to the fact that the last two
references do not include $\Delta b_{\rm I}$, which leads to an
apparent suppression of the effect (see Fig.\ref{fig:dbias}). Another
possible source of discrepancy may be choice of the halo finder which,
as seen in Fig.\ref{fig:fnu2},  has an impact on the strength of the
non-Gaussian correction to the  mass function. This possibility is
investigated in Figure~\ref{fig:dbfof}, which shows the non-Gaussian
bias correction obtained with FOF halos. For this low biased sample,
the scale-independent correction is $|\Delta b_{\rm I}|\lesssim 0.003$
and can thus be neglected. The best-fit values of $\floc$ are
significantly  below the input values of $\pm 100$, in agreement with
the findings of \cite{2009MNRAS.398..321G,2010MNRAS.402..191P} (note,
however, that this suppression is more consistent with $\dc$ being
rescaled by $\sqrt{q}\dc\approx 0.86\dc$ and $b_{\rm L}$ being
unchanged).  This indicates that the choice of halo finder may also
affect the magnitude of the scale-dependent non-Gaussian
bias. Discrepancies have also been observed between the theoretical
and measured non-Gaussian bias corrections  in non-Gaussian models of
the local cubic-order coupling $\gloc\phi^3$
\cite{2010PhRvD..81b3006D}. Understanding all these results clearly
requires a better modeling of halo clustering.

\subsubsection{Redshift-space distortions}

Peculiar velocities generate systematic differences between the
spatial distribution of data in real and redshift space. These
redshift-space  distortions must be properly taken into account in
order to extract  $\fnl$ from redshift surveys.  On the linear scales
of interest, the redshift-space power spectrum of biased tracers reads
as \cite{1987MNRAS.227....1K,1998ASSL..231..185H}
\begin{equation}
P_{\rm h}^s(k,\mu)=\Bigl[b_1^2 P_\delta(k)+2 b_1 f\mu^2 
P_{\delta\theta}(k)+f^2\mu^4P_\theta(k)\Bigr]\;,
\end{equation}
where $P_{\delta\theta}$ and $P_\theta$ are the density-velocity and
velocity divergence power spectra, $\mu$ is the cosine of the angle
between the wavemode $\vk$ and the line of sight and $f$ is the
logarithmic derivative of the growth factor. For $P_\theta$, the
1-loop correction due to primordial NG is identical to
Eq.(\ref{eq:p12ng}) provided  $F_2(\vk_1,\vk_2)$ is replaced by the
kernel  $G_2(\vk_1,\vk_2)=3/7+\mu(k_1/k_2+k_2/k_1)/2+4\mu^2/7$
describing the 2nd order evolution of the velocity divergence
\cite{2000ApJ...542....1S}.  For $P_{\delta\theta}$, this correction is
\begin{align}
\Delta P_{\delta\theta}^{\rm NG}(k) & =\int\!\!\frac{d^3\vq}{(2\pi)^3}\,
\Bigl[F_2(\vq,\vk-\vq)+G_2(\vq,\vk-\vq)\Bigr] \nonumber \\
& \qquad \times B_0(-\vk,\vq,\vk-\vq)\;.
\end{align}
Again, causality implies that $G_2(\vk_1,\vk_2)$ vanishes in the limit
$\vk_1=-\vk_2$. For unbiased tracers with $b_1=1$, the  linear Kaiser
relation is thus recovered at large scales $k\lesssim 0.01\hmmpc$
(this is consistent with the analysis of
\cite{2010MNRAS.402.2397L}). For biased tracers, we still expect the
Kaiser formula to be valid, but the distortion parameter $\beta$
should now be equal to  $\beta=f/(b_1+\Delta b_\kappa)$, where $\Delta
b_\kappa(k,\fnl)$ is the  scale-dependent bias induced by the
primordial non-Gaussianity.

\subsubsection{Mitigating cosmic variance and shot-noise}

Because of the finite number of large scale wavemodes accessible to a
survey, any large scale measurement of the power spectrum is limited
by the cosmic (or sampling) variance caused by the random  nature of
the wavemodes. For discrete tracers such as galaxies, the shot noise
is another source of error. Restricting ourselves to weak primordial
NG, the relative error on the power spectrum $P$ is $\sigma_P/P\approx
1/\sqrt{N}(1+\sigma_n^2/P)$, where $N$ is the number of independent
modes measured and $\sigma_n^2$ is the shot-noise
\cite{1994ApJ...426...23F}. Under the standard assumption of Poisson
sampling, $\sigma_n^2$ equals the inverse of the number density
$1/\bar{n}$ and causes  a scale-independent enhancement of the power
spectrum. The extent to which one can improve the observational limits
on the nonlinear parameters $\fnl$ will strongly depend on our ability
to minimize the impact  of these two sources of errors. By comparing
differently biased tracers  of the same surveyed volume
\cite{2009PhRvL.102b1302S,2009JCAP...10..007M} and suitably weighting
galaxies (by the mass of their host halo for instance)
\cite{2009JCAP...03..004S,2009PhRvL.103i1303S}, it should be possible
to circumvent these problems and considerably improve the detection
level.

\begin{figure}
\center \resizebox{0.5\textwidth}{!}{\includegraphics{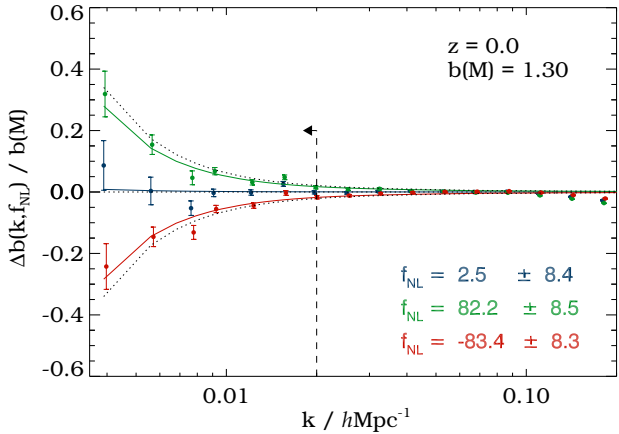}}
\caption{Fractional correction to the Gaussian halo bias in the
$\floc=\pm 100$ and Gaussian models. In contrast to
Fig.~\ref{fig:powng}, halos were identified with a FOF finder of
linking length  $b=0.2$. Only the wavemodes to the left of the
vertical line were used to fit $\Delta b_\kappa(k,\floc)$. For this
low biased sample, the scale-independent correction is  $|\Delta
b_{\rm I}|\lesssim 0.003$ and can thus be ignored. The best-fit value
of $\floc$ and the corresponding 1$\sigma$ error is quoted for each
model (Figure taken from \cite{nicohamaus}).}
\label{fig:dbfof}
\end{figure}

Figure \ref{fig:dbfof} illustrates how the impact of sampling variance
on the measurement of $\floc$ can be mitigated. Namely, the data
points show the result of taking the ratio
$\phh(k,\floc)/\pmm(k,\floc)$  for each set of runs with same Gaussian
random seed field $\phi$ before averaging over the realizations. This
procedure is  equivalent to the multi-tracers method advocated by
\cite{2009PhRvL.102b1302S}. Here, $\pmm$ can be thought as mimicking
the power spectrum of a nearly unbiased tracer of the mass density
field with high number density. Although, in practical applications,
using the dark matter field works better \cite{nicohamaus}, in real
data  $\pmm$ should be replaced by a tracer of the same surveyed
volume  different than the one used to compute $\phh$.  Figure
\ref{fig:dbfof} also shows that, upon taking out most of the cosmic
variance, there is some residual noise caused by the discrete nature
of the dark matter halos. As shown recently \cite{2009PhRvL.103i1303S}
however, weighting the halos according to their mass can dramatically
reduce the shot noise relative to the Poisson  expectation, at least
when compared against the  dark matter. Applying such a weighting may
thus significantly improve the error on the nonlinear parameter
$\floc$, but this should be explored in realistic simulations of
galaxies, especially because the halo mass $M$ may not be easily
measurable from real data \cite{nicohamaus}.  This approach
undoubtedly deserves further attention as it has the potential to
substantially improve the extraction of the primordial non-Gaussian
signal from galaxy surveys.

To conclude this Section, it is worth noting that, while the PDF of
power values $P(\vk)$ has little discriminatory power (for large
surveyed volume, it converges towards the Rayleigh distribution as a
consequence of the central limit theorem) \cite{1995PhRvD..51.6714F},
the covariance of power spectrum measurements (which is sensitive to
the selection function, but also to correlations among the phase of
the Fourier modes) may provide quantitative limits on certain type of
non-Gaussian models \cite{1994ApJ...426...23F,1996MNRAS.283L..99S}.

\subsection{Galaxy bispectrum and higher order statistics}
\label{sub:bispgal}

Higher statistics of biased tracers, such as the galaxy bispectrum,
are of great interest as they are much more sensitive to the shape
of the primordial 3-point function than the power spectrum 
\cite{2004PhRvD..69j3513S,2007PhRvD..76h3004S,2009ApJ...703.1230J,
2009PhRvD..80l3002S,2009arXiv0911.4768N}. Therefore, they could
break some of the degeneracies affecting the non-Gaussian halo
bias (For example, the leading order scale-dependent correction
to the Gaussian bias induced by the local quadratic and cubic
coupling are fully degenerated \cite{2010PhRvD..81b3006D}).

\subsubsection{Normalized cumulants of the galaxy distribution}

The skewness of the galaxy count probability distribution function
could provide constraints on the amount of non-Gaussianity in the
initial conditions. As discussed in \S\ref{sec:matterng} however, it
is difficult to disentangle the primordial and gravitational causes of
skewness in low redshift data unless the initial density field is
strongly non-Gaussian. The first analyzes of galaxy catalogs in terms
of count-in-cells densities all reached the conclusion that the
skewness (and higher-order moments) of the observed galaxy count PDF
is consistent with the value predicted by gravitational instability of
initially Gaussian fluctuations
\cite{1991MNRAS.253..727C,1992ApJ...398L..17G,1993ApJ...403..450G,
1993ApJ...417...36B,1994ApJ...429...36F,1995MNRAS.274.1049B}. Back
then however, most  of the galaxy samples available were not large
enough to accurately determine the $S_J$ at large scales
\cite{1995ApJ...443..469L}. Despite the two orders of magnitude
increase in surveyed volume, these measurements are still sensitive to
cosmic variance, i.e. to the presence of massive super-clusters or
large voids. Nevertheless, the best estimates of the first normalized
cumulants $S_J$ of the galaxy PDF strongly suggest that high order
galaxy correlation functions indeed follow the hierarchical scaling
predicted by the gravitational clustering of Gaussian ICs
\cite{2004MNRAS.352.1232C}. There is no evidence for strong
non-Gaussianity in the initial density field as might by seeded by
cosmic strings or textures \cite{2006MNRAS.373..759F}.

The genus statistics of constant density surfaces through the  galaxy
distribution measures the relative abundance of low and high density
regions as a function of the smoothing scale $R$  and, therefore,
could also be used as a diagnostic tool for primordial
non-Gaussianity. For a Gaussian random field, the genus curve
(i.e. the genus number as a function of the density contrast) is
symmetric about $\delta_R=0$ regardless the value of $R$.  Primordial
NG and nonlinear gravitational evolution can disrupt this symmetry
\cite{1996ApJ...460...51M}. The effect of non-Gaussian ICs on the
topology of the galaxy distribution has been explored in a number of
papers
\cite{1992MNRAS.259..652W,1993MNRAS.260..572C,1996ApJ...463..409M,
2001ApJ...556..641H,2008MNRAS.385.1613H}. For large values of $R$ and
realistic amount of primordial NG, the genus statistics can also be
expanded in a series whose coefficients are the normalized cumulants
$S_J$ of the smoothed galaxy density field. In other words, the genus
statistics essentially  provides another measure of the (large scale)
cumulants. So far,  measurements from galaxy data are broadly
consistent with Gaussian initial conditions
\cite{2008ApJ...675...16G,2009MNRAS.394..454J}.

\subsubsection{Galaxy bispectrum}

Most of the scale-dependence of the primordial $n$-point functions is 
integrated out in the normalized cumulants, which makes them weakly
sensitive to primordial NG. However, while the effect of non-Gaussian
initial conditions, galaxy bias, gravitational instabilities etc. are 
strongly degenerated in the $S_J$, they imprint distinct signatures
in the galaxy bispectrum $B_g(\vk_1,\vk_2,\vk_3)$, an accurate
measurement of which could thus constrain the shape of the primordial
3-point function. 

In the original derivation of \cite{2007PhRvD..76h3004S}, the large
scale (unfiltered) galaxy bispectrum in the $\floc$ model is given by
\begin{align}
\label{eq:bispg}
B_g(\vk_1,\vk_2,\vk_3)&= b_1^3 B_0(\vk_1,\vk_2,\vk_3) \\
& \quad +b_1^2 b_2\Bigl[P_0(k_1) P_0(k_2)+\mbox{(cyc.)}\Bigr] 
\nonumber \\
& \quad + 2b_1^3\Bigl[F_2(\vk_1,\vk_2)P_0(k_1)P_0(k_2)
+\mbox{(cyc.)}\Bigr] \nonumber \;.
\end{align}
Again, $b_1$ and $b_2$ are the first- and second-order bias parameters
that describe the galaxy biasing relation assumed local and
deterministic \cite{1993ApJ...413..447F}. The first term in  the
right-hand side is the primordial contribution which, for  equilateral
configurations and in the $\floc$ model, scales as $\TM(k,z)^{-1}$
like in the matter bispectrum, Eq.(\ref{eq:bispm}). The two last terms
are the contribution from nonlinear bias and the tree-level correction
from gravitational instabilities, respectively. They have the smallest
signal in squeezed configurations.

As recognized by \cite{2009ApJ...703.1230J,2009PhRvD..80l3002S},
Eq.(\ref{eq:bispg}) misses an important term that may significantly
enhance the sensitivity of the galaxy bispectrum to non-Gaussian
initial conditions. This contribution is sourced by the trispectrum
$T_R(\vk_1,\vk_2,\vk_3,\vk_4)$ of the smoothed mass density field,
\begin{equation}
\frac{1}{2}b_1^2 b_2\int\!\!\frac{d^3 q}{(2\pi)^3}\,
T_R(\vk_1,\vk_2,\vq,\vk_3-\vq)+\mbox{(2 perms.)}\;.
\end{equation}
At large scale, this simplifies  to the sum of the linearly evolved
primordial trispectrum $T_0(\vk_1,\vk_2,\vk_3,\vk_4)$ and a coupling
between the primordial bispectrum $B_0(\vk_1,\vk_2,\vk_3)$  (linear in
$\fnl$) and the second order PT corrections (through the kernel
$F_2(\vk_1,\vk_2)$). In the case of local non-Gaussianity and for
equilateral configurations, the first piece proportional to $T_0$
scales as $(\floc)^2 k^{-4}$ times the Gaussian tree-level prediction,
with the same redshift dependence. Hence, it  is similar to the second
order correction $(\floc)^2{\cal M}_R^{-2}P_R(k)$ that appears in the
halo power spectrum (see Eq.\ref{eq:phhloc}). The second piece linear
in $\fnl$ generates a signal at large scales  for essentially all
triangle shapes in the local model as well as in the case of
equilateral NG.  This second contribution is maximized in the squeezed
limit (where it is one order of magnitude larger than the result
obtained by \cite{2007PhRvD..76h3004S}) which helps disentangling it
from the Gaussian terms. Note that a strong dependence on triangle
shape is also present in other NG scenarios such as the $\chi^2$ model
\cite{2000ApJ...542....1S}.

This newly derived contributions are claimed to lead to more than one
order of magnitude improvement in certain limits
\cite{2009ApJ...703.1230J}, but it is not yet clear whether these
gains can be fully realized with upcoming galaxy surveys.  To
accurately predict the constraints that could be achieved with future
measurements of the galaxy bispectrum, a comparison of these
predictions with the halo bispectrum extracted from numerical
simulations is highly desirable. To date, the only numerical study
\cite{2009arXiv0911.4768N} has measured the halo bispectrum for some
isosceles triangles ($k_1=k_2$). While the shape dependence is in
reasonable agreement with the theory, the  observed $k$-dependence
appears to depart from the predicted scaling.

\subsection{Intergalactic medium and the Ly$\alpha$ forest}

\begin{figure}
\center
\resizebox{0.5\textwidth}{!}{\includegraphics[angle=-90]{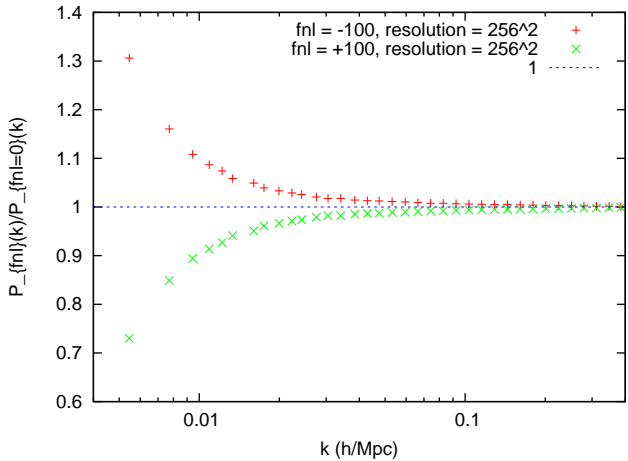}}
\caption{Ratio between the $z=2$ Ly$\alpha$ flux power spectrum extracted
from simulations of Gaussian and non-Gaussian initial conditions. The 
mean transmission is set to $\bar{F}=0.8$ and the power-law exponent
$\alpha=1.65$ (see text).}
\label{fig:flux}
\end{figure}

Primordial non-Gaussianity also affects the intergalactic medium (IGM)
as a positive $\fnl$ enhances the formation of high-mass halos  at
early times and, therefore, accelerates reionization
\cite{2003MNRAS.346L..31C,2006MNRAS.371.1755A,2009MNRAS.394..133C}.
At lower redshift, small box hydrodynamical simulations of the
Ly$\alpha$ forest indicate  that non-Gaussian initial conditions could
leave a detectable signature in the Ly$\alpha$ flux PDF, power
spectrum and bispectrum  \cite{2009MNRAS.393..774V}. However, while
differences appear quite pronounced in the high transmissivity tail of
the flux PDF (i.e. in underdense regions), the Ly$\alpha$ 1D flux
power spectrum seems little affected.  Given the small box size of
these hydrodynamical simulations,  it is worth exploring the effect in
large N-body cosmological simulations using a semi-analytic modeling
of the Ly$\alpha$ forest \cite{shirleyho}, even though such an
approach only provides a very crude approximation to the
temperature-density diagram of the IGM in hydrodynamical simulations.
Figure ~\ref{fig:flux} shows the imprint of local type NG on the
Ly$\alpha$ 3D flux power spectrum (which is not affected by projection
effects) extracted at $z=2$ from a series of large simulations. The
Ly$\alpha$ transmitted flux is calculated in the Gunn-Peterson
approximation \cite{1965ApJ...142.1633G}.  A clear signature similar
to the non-Gaussian halo bias can be seen. As expected, it is of
opposite sign since the Ly$\alpha$ forest is anti-biased relative to
the mass density field (overdensities are mapped onto relatively low
flux transmission).

To estimate the strength of the signal (see \cite{shirleyho} for the
details), one can assume that the (real space) optical depth
$\tau(\vx)$ to Ly$\alpha$ absorption at comoving position $\vx$ is
approximately \cite{1998ApJ...495...44C}
\begin{equation}
\tau(\vx)=\bar{\tau}\Bigl[1+\delta_{\rm g}(\vx)\Bigr]^\alpha\;,
\label{optical}
\end{equation}
where $\delta_{\rm g}$ is the gas density, $\bar{\tau}\sim 1$ is the
optical depth at mean gas density and $\alpha\sim 1-2$ is some
parameter  that depends on the exact thermal history of the low
density IGM. The above relation holds for the moderate overdensities
$\delta_{\rm g}\lesssim 10$ that are responsible for most of the
Ly$\alpha$ absorption features. To relate the gas density to the
smoothed linear density field, we could make the simple ansatz
$\delta_{\rm g}\equiv\delta_R$ \cite{2003ApJ...590....1Z}. In this
linear approximation however, the large scale bias $b_{\rm F}$ of the
Ly$\alpha$ flux density field is much larger than that measured in
detailed numerical simulations (e.g., $b_{\rm F}^2\simeq 0.017$ at
$z=3$ \cite{2003ApJ...585...34M}). Therefore, one may want to consider
the lognormal mapping \cite{1991MNRAS.248....1C,1997ApJ...479..523B}
\begin{equation}
\label{eq:gasd}
1+\delta_{\rm g}=\exp(\delta_R-\sigma_R^2/2)
\end{equation}
to better capture nonlinearities in the gas density field. Expanding
$\exp(\delta_R)$ at second order and noticing that,  in the presence
of weak non-Gaussianity, the joint PDF
$P(\delta_R(\vx_1),\delta_R(\vx_2))$ can generically be expanded into
an Edgeworth series where the primordial 3-point function is the
dominant correction, it is straightforward to compute the Ly$\alpha$
3D flux power spectrum for nonzero $\fnl$. Upon a Fourier
transformation, we arrive at
\begin{equation}
\frac{P_{\rm F}(k,\fnl)}{P_{\rm F}(k,0)} = 1 - 4 g_{\rm F}
\sigma_R^2\TM(k,z)^{-1}{\cal F}(k,\fnl)\;.
\end{equation}
where $g_{\rm F}$ is some auxiliary function of
$(\bar{\tau},\alpha,\sigma_R)$.   This result is valid for any model
of primordial NG characterized by an initial bispectrum. In the
$\floc$ model, the large-scale non-Gaussian Ly$\alpha$ bias scales as
$\Delta b_{\rm F}(k,\fnl)\approx -2g_{\rm F}\sigma_R^2\TM(k,z)^{-1}
\propto k^{-2} T(k)^{-1}$ like the non-Gaussian halo bias.  Assuming
$\bar{\tau}=0.7$, $\sigma_R=1.8$ and $\alpha=1.65$ yields a mean flux
$\bar{F}\approx 0.8$  and a ratio $P_{\rm F}(k=0.01,\fnl)/P_{\rm
F}(k=0.01,0)\approx 1\mp 0.13$ for $\floc=\pm 100$ comparable in
magnitude to that seen in Fig.~\ref{fig:flux}. A detection of this
effect, although challenging in particular because of continuum
uncertainties, could be feasible with future data sets.  Summarizing,
the Ly$\alpha$ should provide interesting information on the
non-Gaussian signal over a range of scale and redshift not easily
accessible to galaxy and CMB observations
\cite{2009MNRAS.393..774V,shirleyho}.

\section{Current limits and prospects}
\label{sec:limits}

As the importance of primordial non-Gaussianity relative to the
non-Gaussianity induced by gravitational clustering and galaxy bias
increases towards high redshift, the optimal strategy to constrain 
the nonlinear coupling parameter(s) with LSS is to use large scale,
high-redshift observations \cite{2001MNRAS.325..412V}.

\subsection{Existing constraints on primordial NG}

The non-Gaussian halo bias presently is the only LSS method that
provides a robust limit on the magnitude of a primordial 3-point
function of the local shape. It is a broadband effect that can be
easily measured with photometric redshifts. The authors of
\cite{2008JCAP...08..031S} have applied Eq.(\ref{eq:dbiask}) to
constrain the value of $\floc$ using a compilation of large-scale
clustering data. Their constraint arise mostly from the QSO sample  at
median redshift $z=1.8$, which covers a large comoving volume  and is
highly biased, $b_1=2.7$. They obtain
\begin{equation}
-29 < \floc < +69
\end{equation}
at 95\% confidence level.  These limits are competitive with those
from CMB measurements,  $-10<\floc<+74$ \cite{2010arXiv1001.4538K}.
It is straightforward  to translate this  2-$\sigma$ limit into a
constraint on the cubic order coupling $\gloc$ since the non-Gaussian
scale-dependent bias  $\Delta b_\kappa(k,\gloc)$ has the same
functional form as  $\Delta b_\kappa(k,\floc)$
\cite{2010PhRvD..81b3006D}. Assuming $\floc=0$, one  obtains
\begin{equation}
-3.5\times 10^5 < \gloc < +8.2 \times 10^5\;.
\end{equation}
These limits are comparable with those inferred from the analysis of
CMB data.

Measurements of the galaxy bispectrum in several redshift catalogs
have shown evidence for a configuration shape dependence in agreement
with that predicted from gravitational instability, ruling out
$\chi^2$ initial conditions at the 95\%
C.L. \cite{1999ApJ...521L..83F,2001ApJ...546..652S}.  Recent analyses
of the SDSS LRGs catalogue indicate that the shape dependence  of the
reduced 3-point correlation $Q_3\sim\xi_3/(\xi_2)^2$ is  also
consistent with Gaussian ICs  \cite{2007MNRAS.378.1196K}, although  a
primordial (hierarchical)  non-Gaussian contribution in the range
$Q_3\sim 0.5-3$ cannot be ruled out  \cite{2009MNRAS.399..801G}. Other
LSS probes of primordial non-Gaussianity, such as the abundance of
massive clusters, are still too affected by systematics to furnish
tight constraints on the shape and magnitude of a primordial 3-point
function. Still, the observation of a handful of unexpectedly massive
high-redshift clusters has been interpreted as evidence of a
substantial  degree of primordial NG
\cite{2000ApJ...530...80W,2009PhRvD..80l7302J, 2010arXiv1003.0841S}.

\subsection{Future prospects}

Improving the current limits will further constrain the physical 
mechanisms for the generation of cosmological perturbations. 

The non-Gaussian halo bias also leaves a signature in
cross-correlation  statistics of weak cosmic shear (galaxy-galaxy and
galaxy-CMB) \cite{2009PhRvD..80l3527J,2009arXiv0912.4112F} and in the
integrated Sachs-Wolfe (ISW) effect
\cite{2008JCAP...08..031S,2008PhRvD..78l3507A,2008ApJ...684L...1C}.
Measurements of the lensing bispectrum could also  constrain a number
of non-Gaussian models \cite{2004MNRAS.348..897T}.   However, galaxy
clustering will undoubtedly offer the most promising LSS diagnostic of
primordial non-Gaussianity. The detectability of a local primordial
bispectrum has been assessed in a series of papers. It is expected
that future all-sky galaxy surveys will achieve constraints of the
order of $\Delta\floc\sim 1$ assuming all systematics are reasonably
under control
\cite{2008JCAP...04..014L,2008JCAP...08..031S,2008PhRvD..78l3507A,
2008PhRvD..78l3519M,2008ApJ...684L...1C,2010RAA....10..107G,
2009JCAP...12..022S,2010arXiv1003.0456C}.  Realistic models of cubic
type non-Gaussianity \cite{2010PhRvD..81b3006D}, modifications of the
initial vacuum state or horizon-scale GR corrections
\cite{2009ApJ...706L..91V} should also be tested with future
measurement of the galaxy power spectrum.

Upcoming observations of high redshift clusters will provide increased
leverage on measurement of primordial non-Gaussianity with abundances
and possibly put limits on any nonlinear parameter $\fnl$ at the level
of a few tens \cite{2009MNRAS.397.1125F}.  Combining the information
provided by the evolution of the mass function and power spectrum of
galaxy clusters should yield constraints with a precision
$\Delta\floc\sim 10$ for a wide field survey covering half of the sky
\cite{2010arXiv1003.0841S}. Alternatively, using the full covariance
of cluster counts (which is sensitive to the non-Gaussian halo bias)
can furnish constraints of $\Delta\floc\sim 1-5$ for a Dark Energy
Survey-type experiment \cite{2009PhRvL.102u1301O,2010arXiv1003.2416C}.

As emphasized in \S\ref{sec:lssprobes} however, the exact magnitude of
the non-Gaussian bias is still uncertain partly due to the freedom at
the definition of the halo mass and the uncertainty in the
correspondence between simulated quantities and observables.
Understanding this type of systematics will be crucial to set reliable
constraints on a primordial non-Gaussian component.  To fully exploit
the potential of future galaxy surveys, it will also be essential to
extend the theoretical and numerical analyses to other bispectrum
shapes than the local template used so far. Ultimately, the gain that
can  be achieved will critically depend on our ability to minimize the
impact of sampling variance and shot-noise. In this regards,
multi-tracers methods combined with optimal weighting schemes should
deserve further  attention as they hold the promise to become the most
accurate method to extract the primordial non-Gaussian signal from
galaxy surveys \cite{2009PhRvL.102b1302S,2009JCAP...03..004S,
2009JCAP...10..007M,2009PhRvL.103i1303S}.

\section{Acknowledgments}

We give our special thanks to Nico Hamaus and Shirley Ho for sharing
with us material prior to publication, and Emiliano Sefusatti for
providing us with the data shown in Fig.\ref{fig:bisp}. We would also
like  to thank Martin Crocce, Christopher Hirata, Ilian Iliev, Tsz Yan
Lam, Patrick  McDonald, Nikhil Padmanabhan, and Anze Slosar  for
collaboration on these issues, and Tobias Baldauf for comments and for
a careful reading of this manuscript.  This work was supported by the
Swiss National Foundation (Contract  No. 200021-116696/1) and made
extensive use of the NASA Astrophysics  Data System and and arXiv.org
preprint server.

\bibliographystyle{unsrt}
\bibliography{aa_reviewng}

\end{document}